\documentclass[a4paper,11pt]{article}

\usepackage[utf8]{inputenc}
\usepackage[T1]{fontenc}
\usepackage[english]{babel}

\usepackage{indentfirst}
\usepackage{amsmath}
\usepackage{amsthm}
\usepackage{amssymb}
\usepackage{graphicx}
\usepackage[font=small]{caption}
\usepackage{subcaption}
\usepackage{psfrag}
\usepackage{color}
\usepackage{pgf,pgfnodes}

\allowhyphens

\newcommand{\fa}{\mathfrak{a}}

\newcommand{\valos}{\mathbb{R}}
\newcommand{\complex}{\mathbb{C}}

\newcommand{\ordo}{\mathcal{O}}
\newcommand{\Pe}{{P}}

\newcommand{\CC}{\mathcal{C}}

\newcommand{\ket}[1]{{\left|#1\right\rangle}}
\newcommand{\bra}[1]{{\left\langle #1\right|}}

\newcommand{\skalarszorzat}[2]{{\langle #1 | #2 \rangle}}

\newcommand{\vev}[1]{\left\langle #1 \right\rangle}

\newtheorem{conj}{Conjecture}

\usepackage{ifpdf}

\ifpdf
\usepackage{epstopdf}
\usepackage[pdftex,colorlinks,urlcolor=blue,citecolor=blue,linkcolor=blue]{hyperref}
\else
\usepackage[hypertex,colorlinks,urlcolor=blue,citecolor=blue,linkcolor=blue]{hyperref}
\fi

\makeatother

\begin{document}

\numberwithin{equation}{section}

\title{Excited state correlations of the finite Heisenberg chain}
\author{Bal\'azs Pozsgay$^1$
\\
~\\
$^{1}$MTA--BME \textquotedbl{}Momentum\textquotedbl{} Statistical
Field Theory Research Group\\
Budafoki \'ut 8, H-1111 Budapest, Hungary\\
}
\maketitle

\abstract{
We consider short range correlations in excited states of the
finite XXZ and XXX Heisenberg spin chains. 
We conjecture that the known results for the factorized ground state
correlations can be applied to the excited states too, if
the so-called physical part of the construction is changed
appropriately. For the ground state we derive simple algebraic expressions
for the physical part; the formulas only use the ground state Bethe roots as an
input. We conjecture that the same formulas can be applied to the
excited states as well, if the exact Bethe roots of the excited states
are used instead.
In the XXZ chain the results are expected to be valid for all
 states (except certain singular cases where regularization is
needed), whereas in the XXX case they only apply to singlet states or
group invariant operators. Our conjectures are tested against numerical
data from exact diagonalization and coordinate Bethe Ansatz
calculations, and perfect agreement is found in all cases. In the XXX
case we also derive a new result for the
  nearest-neighbour correlator $\vev{\sigma_1^z\sigma_2^z}$, which is
  valid for non-singlet states as well.
Our results build a bridge between the known theory of factorized
correlations, and the recently conjectured TBA-like description for
the building blocks of the construction.
}

\section{Introduction}

The Heisenberg spin chain is a model of magnetism in one-dimensional
or quasi one-dimensional materials. The study of the original XXX
model goes back to its
famous solution by Hans Bethe in 
1931 \cite{XXX}, whereas the anisotropic version (also called the XXZ
model) was first solved by Orbach in 1958 \cite{XXZ1}.
These spin chains play a central role in the field of integrable
models: they are truly interacting models whose solution displays the
full arsenal of integrability, yet their relative simplicity make them an
ideal testing ground to develop new ideas and methods. 

By now a large body of literature has been devoted to the study of the
equilibrium properties of the spin chain.
The exact eigenstates can be constructed using various forms of
the Bethe Ansatz \cite{KorepinBook}, and the thermodynamic properties
can be computed using the so-called Thermodynamic Bethe Ansatz (TBA) or the
Quantum Transfer Matrix (QTM) methods
\cite{TakahashiBook,kluemper-QTM}. On the other hand, for a long time it was believed
that the correlation functions can not be computed in a practical
way. The correlators are important physical quantities: they are
experimentally relevant, and a
system can not be considered to be exactly solved until (at least some
of the) correlators can be computed. This motivated a long series
of works by different groups to study the correlators of the
Heisenberg spin chain.

The first results were multiple integral
representations for the ground state correlations, which were derived using representation theory of quantum
algebras \cite{jimbo-miwa1,jimbo-miwa2,Jimbo-Miwa-book} or the Algebraic Bethe Ansatz
\cite{KitanineMailletTerras-XXZ-corr1,QTM1}. Later it was realized in the
papers
\cite{boos-korepin-first-factorization,boos-korepin-smirnov-fact2,boos-korepin-smirnov-fact3,bjmst-fact4}
that for the ground state of the infinite XXX model these 
multiple integrals can be factorized, i.e. expressed as polynomials of
a single function in two variables. An exponential formula was found
in \cite{bjmst-fact5,bjmst-fact6} for the reduced density matrix of a finite sub-chain, whose
form was conjectured to be valid even in the finite temperature or
finite length cases
\cite{boos-goehmann-kluemper-suzuki-fact7,goehmann-kluemper-finite-chain-correlations}.
Afterwards a new fermionic structure was found on the space of local
operators of the XXZ model \cite{hgs1,hgs2,HGSIII}, which led to
easily manageable expressions for the short range correlators
including the finite temperature or finite length cases
\cite{physpart,XXZ-massless-corr-numerics-Goehmann-Kluemper,XXZ-massive-corr-numerics-Goehmann-Kluemper,kluemper-goehmann-finiteT-review}. In 
practical terms these developments can be summarized as follows: In both the XXX and
XXZ cases the correlations can be expressed as a polynomial of only
one or two functions, respectively\footnote{This applies to
  spin-reflection invariant operators. In the generic case
  (including for example the magnetization operator) one more
function is needed.}. The algebraic part of the
construction provides this polynomial, whereas the physical part specifies
the functions themselves depending on the physical situation, which
might be the infinite chain at finite temperature, or the ground state
of the finite chain.
We should also note, that an
independent derivation of these results was given later in
\cite{kluemper-discrete-functional} using discrete functional equations.

The previously mentioned results pertain to equilibrium
situations. However, recently there has been considerable interest in
the far from equilibrium physics of integrable models, including and
especially the Heisenberg spin chains
\cite{essler-fagotti-quench-review,rigol-quench-review}. One of the
main questions was whether an integrable model equilibrates to some
kind of Generalized Gibbs Ensemble \cite{rigol-gge,rigol-quench-review}. Regarding the
Heisenberg chain this question has been investigated in a series of
works
\cite{sajat-xxz-gge,essler-xxz-gge,fagotti-collura-essler-calabrese,JS-oTBA,sajat-oTBA}
leading to \cite{JS-CGGE} (see also \cite{jacopo-michael-hirota}), where a conclusive answer was given: the
asymptotic states can indeed be described by a generalized statistical physical
ensemble, if the recently discovered quasi-local charges are also
included \cite{Prosen1,Prosen2,kvazilok2,prosen-xxx-quasi}. However,
the addition of all charges completely fixes all the string root densities
of the spin chain  \cite{JS-CGGE,essler-fagotti-quench-review}, therefore it is a question of
interpretation whether there is any kind of statistical physics
emerging in the long time limit. 

In all of these studies it was of central importance to give predictions for the long-time
limit of local observables, so that the analytic results could be
compared to independent simulations \cite{JS-oTBA,sajat-oTBA} or
possibly to experimental data. In out-of-equilibrium situations the
system is typically very far from the ground state and there is need
to calculate the correlations in highly excited states too. In the
spin chain literature the first such results were presented in
\cite{sajat-corr}, where it was conjectured that in the thermodynamic
limit the known factorized formulas are valid even for the highly
excited states if the physical part
is computed using a new set of TBA-like integral equations. This
conjectured result was used in the works \cite{JS-oTBA,sajat-oTBA}, it
successfully passed a number of tests, however it was not clear how it
relates to the physical part of the finite temperature situation
\cite{HGSIII,physpart}. In the latter problem all
ingredients are computed using single contour integrals, whereas \cite{sajat-corr} uses an
infinite system of equations based on the string hypothesis. It is
important that the results of \cite{sajat-corr}  are valid for arbitrary smooth string
distributions, and not only for the thermal cases. 
For the the free energy of the finite $T$ case it is known how to
connect the TBA equations
 to the single non-linear integral equation of the QTM method \cite{TBA-QTM-Kluemper-Takahashi}, but
up to now no such link was known for the factorized correlation functions.

Here we make a step towards filling this gap by
investigating the correlations of the excited states of the
finite spin chain; this problem has not yet been considered in the
literature. We derive algebraic expressions for the physical part of
the factorized correlation functions; the results are expected to
be valid for all excited states. In the thermodynamic limit these
results could lead to a proof of the formulas of \cite{sajat-corr}.

The structure of this article is as follows. In Section \ref{sec:XXZ}
we present one of the main conjectures, namely that factorization holds for all
excited states of the XXZ model. Also, we present a formula for the
physical part, which is a simple algebraic expression that uses the exact
Bethe roots. Section \ref{sec:XXX} deals with the correlations of the
XXX model; the focus is on singlet states and singlet operators.
In Section \ref{sec:HF} we derive a simple but new result 
for the nearest neighbour $z-z$ correlator of the XXX chain, which is
valid for arbitrary Bethe states and not only for the
singlets. Section \ref{sec:CO} includes our conclusions, and also an
outlook to open problems. Finally, Appendices \ref{sec:numerics} and \ref{sec:app1} include numerical
data and simple coordinate space calculations to support our results,
whereas in Appendix \ref{sec:appzz} we compare a result of the paper \cite{boos-goehmann-kluemper-suzuki-fact7} to
one of our finite size formulas.

\section{Excited state correlations of the XXZ model}

\label{sec:XXZ}

In this section we consider the homogeneous XXZ spin chain for generic anisotropy. The
model is defined by the Hamiltonian
\begin{equation}
\label{H}
  H=\sum_{j=1}^L (\sigma_j^x\sigma_{j+1}^x+\sigma_j^y\sigma_{j+1}^y+\Delta(\sigma_j^z\sigma_{j+1}^z-1)).
\end{equation}
In this work we only consider periodic boundary conditions and assume
that $L$ is even. For the anisotropy we will use the parametrization $\Delta=\cosh(\eta)$.

As usual we introduce the monodromy matrix as
\begin{equation}
\label{Tdef}
T(u)=R_{L0}(u)\dots R_{10}(u),
\end{equation}
where the $R_{j0}(u)$ operators are acting on the quantum space at site $j$
and an auxiliary space denoted by 0. 
The matrix elements of $R_{j0}(u)$ are identical to the well known
R-matrix of the XXZ type:
\begin{equation}
  R(u)=\frac{1}{\sin(u+i\eta)}
  \begin{pmatrix}
    \sin(u+i\eta) & & &\\
& \sin(u)  & \sin(i\eta) & \\
& \sin(i\eta) & \sin(u) & \\
& & & \sin(u+i\eta)
  \end{pmatrix}.
\label{R}
\end{equation}
The trace of the monodromy matrix is called the transfer matrix:
\begin{equation*}
  \tau(u)=A(u)+D(u).
\end{equation*}
The transfer matrices form a commuting family:
\begin{equation*}
  [\tau(u),\tau(v)]=0.
\end{equation*}
It is also known that the Hamiltonian and the higher charges of the
model can be obtained as the logarithmic derivative of the transfer
matrix around $u=0$:
\begin{equation*}
 Q_j= \left.\left(\frac{d}{du}\right)^{j+1}\log \tau(iu)\right|_{u=0}.
\end{equation*}
It can be shown that with these conventions $H=2\sinh(\eta)Q_0$.

Eigenstates of the model can be constructed by various forms of the
Bethe Ansatz. The coordinate Bethe Ansatz solution can be written as
follows. We define $N$-particle states as
\begin{equation}
\label{XXZ-eloallitas}
  \ket{\{u\}_N}=
\sum_{y_1< y_2<\dots< y_N}
  \phi_N(\{u\}_N|y_1,\dots,y_N)   \sigma^-_{y_1}\dots \sigma^-_{y_N} \ket{0},
\end{equation}
where $\ket{0}$ is the reference state with all spins up. Then
the wave functions can be written as
\begin{equation}
\label{xxz-coo-ba}
  \phi_N(\{u\}_N|\{y\})=\sum_{\Pe \in S_N} 
\left[\prod_{1\le m < n \le N}\frac{\sin(u_{P_m}-u_{P_n}+i\eta)}{\sin(u_{P_m}-u_{P_n})}\right]
\left[\prod_{l=1}^N
\left(\frac{\sin(u_{P_l}+i\eta/2)}{\sin(u_{P_l}-i\eta/2)}\right)^{y_l}\right].
\end{equation}
Here $u_j$ are the rapidities of the interacting spin waves and
they satisfy the Bethe equations, which follow from the periodicity of
the wave function:
\begin{equation}
 \label{XXZBE}
\left(
\frac{\sin(u_j-i\eta/2)}{\sin(u_j+i\eta/2)}
\right)^L \prod_{k\ne j}  
\frac{\sin(u_j-u_k+i\eta)}{\sin(u_j-u_k-i\eta)}=1.
\end{equation}
The energy eigenvalues are
\begin{equation}
\label{XXZ-energy}
  E=-\sum_j  
\frac{2\sinh^2\eta}{\sin(u_j+i\eta/2)\sin(u_j-i\eta/2)}.
\end{equation}
In the regime $\Delta>1$ we have $\eta\in\valos$ and the solutions to
the Bethe equations \eqref{XXZBE} are either real or they form strings that are
centered at the real axis. On the other hand, for $\Delta<1$ the
parameter $\eta$ is purely imaginary, and as an effect we have a rotation in
the complex plane: if we use the same formulas \eqref{xxz-coo-ba}-\eqref{XXZBE} even in this regime, then
the Bethe roots are either on
the imaginary axis or they form strings centered around
it\footnote{The so-called ``negative-parity'' strings with $\mathcal{R}u=\pi/2$ can also
be  considered to be centered around the imaginary axis due to the
  $\pi$-periodicity.}. Usually 
an explicit rotation is performed 
for $|\Delta|<1$  by using hyperbolic
functions instead of the trigonometric ones. However, in the present
work we intend to treat the two regimes together, therefore we use the
trigonometric formulas for arbitrary $\Delta\ne 1$.

With the convention \eqref{xxz-coo-ba} the norm of the
state \eqref{XXZ-eloallitas} is given by \cite{korepin-norms}
\begin{equation}
\label{norm}
  \skalarszorzat{\{u\}_N }{ \{u\}_N}=\prod_j \frac{\sin(u_j+i\eta/2)\sin(u_j-i\eta/2)}{\sinh(\eta)}
\prod_{j<k} \frac{\sin(u_{jk}+i\eta)\sin(u_{jk}-i\eta)}{\sin^2(u_{jk})} 
\times \det G,
\end{equation}
where $G$ is the Gaudin matrix:
\begin{equation}
\label{XXZG}
  G_{jk}=\delta_{jk}\left(
L\frac{\sinh(\eta)}{\sin(u_j+i\eta/2)\sin(u_j-i\eta/2)} +\sum_{l=1}^N K(u_{jl})
\right)-K(u_{jk}),
\end{equation}
with $u_{jk}=u_j-u_k$ and $K$ is the scattering kernel of the XXZ model:
\begin{equation}
K(u)=-\frac{\sinh(2\eta)}{\sin(u+i\eta)\sin(u-i\eta)}.
\end{equation}
We stress that \eqref{norm} is only valid when the rapidities satisfy
the Bethe equations. In the non-physical off-shell cases the norm is a more complicated
 function of the variables  $\{u\}$.

It is important that even though the Bethe Ansatz seems to be
complete, the regular solutions of the Bethe equations 
\eqref{XXZBE} do not produce all eigenstates of the
XXZ chain \cite{baxter-completeness}. For example, for arbitrary
$\Delta$ there are singular states whose Bethe roots 
include the rapidities $u=\pm i\eta/2$  
\cite{xxx-except-sol,XXX-fine-structure,origin-of-singular-xxx,nepo-singular-aba,XXXmegoldasok,twisting-singular}.
The existence of these states is related to a special property of the
Bethe Ansatz wave function \eqref{xxz-coo-ba}: if the state is an
eigenvector of the space reflection operator, then the corresponding
eigenvalue is always equal to the eigenvalue of the one-site shift
operator, whereas there must be states where these two eigenvalues are
different and this is produced by the singular rapidities \cite{origin-of-singular-xxx}.
Other types of singular states appear at the ``root of unity points''
$\Delta=\cos(\gamma\pi)$ with $\gamma=p/q$ and $p,q\in\mathbb{Z}$
being relative primes;  these states are related to additional
degeneracies in the spectrum caused by the $sl_2$ loop algebra
\cite{sl2rootofunity,fabriziusBetheRoots1,fabriziusBetheRoots2,baxter-completeness}. 
In the present work we concentrate on the regular states and give 
only a few remarks about the singular cases.

\bigskip

Regarding the correlations our focus is on the short range operators, for example 
\begin{equation}
\ordo= E_1^{ab}E_n^{cd},
\end{equation}
where $E_j^{ab}$ is the $\complex^2\to\complex^2$ elementary matrix acting on site $j$ with a
single nonzero matrix element at position $(a,b)$.
Our aim is to compute the excited state mean values
\begin{equation}
  \bra{ \{u\}_N   }\ordo\ket{ \{u\}_N  }.
\end{equation}
The explicit expression for the wave function gives a direct way to compute
the correlators in arbitrary Bethe states. For example
\begin{equation}
  \begin{split}
  &  \bra{ \{u\}_N   } E_1^{22}E_2^{22} \ket{ \{u\}_N  }  =\\
&\hspace{1cm}\sum_{3\le y_2<y_3<\dots<y_N\le L}
 \phi_N^*(\{u\}_N|\{1,2,y_2,\dots,y_N\})
 \phi_N(\{u\}_N|\{1,2,y_2,\dots,y_N\}).
   \end{split}
\end{equation}
In order to study the analytic properties of the correlators it is
useful to introduce the parameters
\begin{equation}
\label{ajdef}
 a_j=e^{ip_j}=\frac{\sin(u_j+i\eta/2)}{\sin(u_j-i\eta/2)}.
\end{equation}
Here $p_j$ can be identified as the one-particle pseudo-momentum and
$a_j$ is the one-particle eigenvalue of the one-site translation
operator. In terms of the $a$-variables the wave function can be written
as 
\begin{equation}
\label{coo-baaa}
  \phi_N(\{a\}_N|\{y\})=\sum_{\Pe \in S_N} 
\left[\prod_{1\le m < n \le N}
\frac{1-2\Delta a_{P_m}+a_{P_m}a_{P_n}}{a_{P_n}-a_{P_m}}
\right]
\left[\prod_{l=1}^N  a_{P_l}^{y_l}
\right].
\end{equation}
We are interested in correlations in the physical states.
The solutions to the Bethe equations are
self-conjugate \cite{Bethe-selfconjugate}, therefore, in terms of the
$a$-variables the conjugate wave function can be written as
\begin{equation}
\label{coo-baaac}
  \phi^*_N(\{a\}_N|\{y\})=\sum_{\Pe \in S_N} 
\left[\prod_{1\le m < n \le N}
\frac{1-2\Delta a_{P_n}+a_{P_m}a_{P_n}}{a_{P_m}-a_{P_n}}
\right]
\left[\prod_{l=1}^N  a_{P_l}^{-y_l}
\right].
\end{equation}
The direct real space calculations lead to expressions that contain
powers of $a_j^L$, $j=1\dots N$. After substituting the 
Bethe equations in the form
\begin{equation*}
  a_j^L =\prod_{k\ne j} -\frac{1-2\Delta a_j+a_ja_k}{1-2\Delta a_k+a_ja_k}
\end{equation*}
all normalized correlators can be written as
\begin{equation}
\label{hatez}
  \bra{\{a\}}\ordo\ket{\{a\}}=\frac{\sum_{j=1}^N L^j C_j(\{a\})}{\sum_{j=1}^N L^j D_j(\{a\})},
\end{equation}
where $C_j$ and $D_j$ are polynomials that don't depend on the volume
$L$ anymore. The denominator in \eqref{hatez} is proportional to the Gaudin
determinant, whereas the polynomials $C_j$ are related to the infinite
volume form factors of the operator in question \cite{sajat-LM}.

The real space calculations are of course cumbersome and it is not clear how to get useful formulas for
arbitrary $N$ and $L$. An alternative and well established method is 
 the Algebraic Bethe Ansatz (ABA), which provides a systematic way towards
 the correlators, 
see \cite{Maillet-xxz-ground-state-corr-review} and references
therein. Previous works concentrated mostly on the ground states, both
at zero and finite magnetic fields, with the aim of taking the
thermodynamic limit. However, they also
include a number of intermediate results for finite
chains involving the Bethe roots as arbitrary
parameters \cite{goehmann-kluemper-finite-chain-correlations}, which are valid for the excited states as
well. 

In the ABA the correlators are first obtained in the form of multiple
integrals. Quite remarkably, these multiple integrals can be
factorized, i.e. expressed as a polynomials of simple integrals. In
the following subsection we summarize the known results for the XXZ
chain, following the presentation of
\cite{goehmann-kluemper-finite-chain-correlations,XXZ-finite-T-factorization,XXZ-massive-corr-numerics-Goehmann-Kluemper}.

\subsection{Factorization of  correlation functions}

The construction for the factorized correlation functions consists of two parts: the 
algebraic part, which deals with the space of operators and
expresses their mean values using two functions, and the
physical part, which computes these functions depending on the physical
situation. The calculations are valid both for the finite size ground
state and at finite temperature in the thermodynamic limit.

As a first step we define the auxiliary function $\fa$ through
\begin{equation}
\label{XXZadef}
\begin{split}
\log \fa(u)=&a_0(u)
+  \int_C \frac{d\omega}{2\pi } K(u-\omega) \log(1+\fa(\omega)).
\end{split}
\end{equation}
The source of the integral equation and the contour depend on the physical situation.
Here we only consider the finite volume ground state case, where 
\begin{equation}
  a_0(x)=L\log \frac{\sin(x-i\eta/2)}{\sin(x+i\eta/2)},
\end{equation}
and $C$ is a narrow contour around the segment $[-\pi/2,\pi/2]$ of the
real axis, so that it encircles all Bethe roots. 

We also define two functions $H(x,y)$ and $\tilde H(x,y)$ through the
linear integral equations\footnote{In the literature the
  function $H$ was denoted by $G$. Here we changed the notation to
  avoid confusion with the Gaudin 
matrix.}
\begin{equation}
\label{XXZHdef}
\begin{split}
H(u,x)=&
-q(u,x)
- \int_C \frac{d\omega}{2\pi} 
K(u-\omega)
\frac{H(\omega,x)}{1+\fa(\omega)}
\end{split}
\end{equation}
and
\begin{equation*}
 \tilde H(u,x)=
-\tilde q(u,x)
-
\int_C \frac{d\omega}{2\pi}
\tilde K(u-\omega)
\frac{H(\omega,x)}{1+\fa(\omega)} 
 -
\int_C \frac{d\omega}{2\pi}K(\lambda-\omega)
\frac{\tilde H(\omega,x)}{1+\fa(\omega)},
\end{equation*}
where
\begin{equation*}
 \tilde  K(u)=\frac{\sin(2u) }{\sin(u+i\eta)\sin(u-i\eta)}
\end{equation*}
and
\begin{equation*}
\begin{split}
  q(u,x)&=-i(\cot(u-x-i\eta)-\cot(u-x))\\
  \tilde q(u,x)&=-i\cot(u-x-i\eta).
\end{split}
\end{equation*}
In these definitions it is assumed that the parameter $x$ lies
within the contour $C$, and in all other cases an analytic 
continuation is understood. This requirement follows from the
derivation of the multiple integrals
\cite{QTM1,goehmann-kluemper-finite-chain-correlations}, where as a
first step an ``inhomogeneous transfer matrix'' has to be
considered. Here we only treat the homogeneous limit. 

For $x,y\in C$ Let the functions $\Psi(x,y)$ and $P(x,y)$ 
be given by
\begin{align}
\label{Psidef}
    \Psi(x,y)&=
\int_C\frac{d\omega}{\pi}
q(\omega,x)
\frac{H(\omega,y)}{1+\fa(\omega)}\\
\label{Pdef}
  P(x,y)&=
\int_C\frac{d\omega}{\pi}\left[
q(\omega,y)
\frac{\tilde H(\omega,x)}{1+\fa(\omega)}
+
\tilde q(\omega,y)
\frac{H(\omega,x)}{1+\fa(\omega)}\right].
\end{align}
The behaviour of these functions in the limits $x,y\to i\eta/2$  or
$x,y\to 0$ determines the correlations in finite
volume or at finite temperature, respectively. Here we are only
interested in the finite volume case, therefore we define 
\begin{equation}
\Psi_{a,b}=\left. \frac{\partial^a}{\partial x^a} 
\frac{\partial^b}{\partial y^b} \Psi(ix,iy)\right|_{x,y=\eta/2},\qquad
P_{a,b}=\left. \frac{\partial^a}{\partial x^a} 
\frac{\partial^b}{\partial y^b} P(ix,iy)\right|_{x,y=\eta/2}.\qquad
\end{equation}
As a final step we define
\begin{equation}
\label{omW}
  \begin{split}
    \omega_{a,b}&=-\Psi_{a,b} -(-1)^b\frac{1}{2}
    \left(\frac{\partial}{\partial u}\right)^{a+b} K(iu)\Big|_{u=0}\\
W_{a,b}&=
-P_{a,b} +(-1)^b\frac{1}{2}
    \left(\frac{\partial}{\partial u}\right)^{a+b} \tilde K(iu)\Big|_{u=0}.
  \end{split}
\end{equation}
The objects $\Psi_{a,b}$ and $\omega_{a,b}$ are symmetric, whereas
$P_{a,b}$ and $W_{a,b}$ are anti-symmetric with respect to the exchange
of indices.

All short distance correlators can be expressed as finite combinations of
the numbers $\omega_{a,b}$ and $W_{a,b}$. Explicit formulas can be found in
the papers
\cite{XXZ-finite-T-factorization,XXZ-massive-corr-numerics-Goehmann-Kluemper}
\footnote{Our notations differ slightly from 
\cite{XXZ-finite-T-factorization,XXZ-massive-corr-numerics-Goehmann-Kluemper}:
The quantities $\omega$ and $W$ correspond to $\omega$ and $\omega'/\eta$ of \cite{XXZ-finite-T-factorization,XXZ-massive-corr-numerics-Goehmann-Kluemper}. 
}.
Simple examples for short range correlators are:
\begin{equation}
\label{corrpeldak}
  \begin{split}
\vev{\sigma^z_1\sigma^z_2}_T&=\coth(\eta)\omega_{0,0}+W_{1,0}\\
\vev{\sigma^x_1\sigma^x_2}_T&=-\frac{\omega_{0,0}}{2\sinh(\eta)}-\frac{\cosh(\eta)}{2} W_{1,0}\\
\vev{\sigma^z_1\sigma^z_3}_T&=2\coth(2\eta)\omega_{0,0}+W_{1,0}+\tanh(\eta)\frac{\omega_{2,0}-2\omega_{1,1}}{4}
-\frac{\sinh^2(\eta)}{4}W_{2,1}\\
\vev{\sigma^x_1\sigma^x_3}_T&=-\frac{1}{\sinh(2\eta)}\omega_{0,0}-\frac{\cosh(2\eta)}{2}
W_{1,0}-\tanh(\eta)\cosh(2\eta)\frac{\omega_{2,0}-2\omega_{1,1}}{8}+\\
&\hspace{6cm} +\sinh^2(\eta)\frac{W_{2,1}}{8}.
  \end{split}
\end{equation}

\subsection{Transforming back to algebraic expressions}

The main idea to get the excited state correlations is to find the
proper modification of the ground state formulas. In the previous
section all the necessary ingredients were presented in the form of contour
integrals. In the field of integrable models it is very common that
the excited state quantities can be obtained by a simple change of
the integration contours; this could be a promising direction
even in our case. In particular, it is plausible that with certain changes
of integration contours all intermediate
results of \cite{goehmann-kluemper-finite-chain-correlations} could be
formulated for the finite volume excited states too, thus leading to
factorized formulas \cite{goehmann-private}.
However, it could be difficult to
define the contours for {\it all} excited states, or to perform
numerical computations in practice. Therefore we choose a different
strategy: we transform the contour integrals into
algebraic expressions, and perform the generalization to excited
states afterwards.

The solution of \eqref{XXZadef} is the well known counting function:
\begin{equation}
\label{asol}
\fa(x)=\left(\frac{\sin(x-i\eta/2)}{\sin(x+i\eta/2)}\right)^L \prod_{k=1}^N  
\frac{\sin(x-u_k+i\eta)}{\sin(x-u_k-i\eta)}.
\end{equation}
The condition $1+\fa(x)=0$ encodes the Bethe equations. Therefore
all integrals involving the weight function $1/(1+\fa(x))$ are naturally
equivalent to a sum over the Bethe roots. For example \eqref{XXZHdef} is
transformed into
\begin{equation}
  H(x,x_1)=-q(x,x_1)-i
\sum_{j=1}^N K(x-u_j) \frac{H(u_j,x_1)}{\fa'(u_j)} 
+ \frac{K(x-x_1)}{1+\fa(x_1)}.
\end{equation}
Here we used the fact that the only pole of $H(x,x_1)$ within the
contour is at $x=x_1$ with residue $i$. For the correlators we will
be interested in the $x_{1,2}\to i\eta/2$ limit (and the first few
derivatives) of $H(x_1,x_2)$. It can be seen from \eqref{asol} that
$\fa(x)$ has an order-$L$ zero at $x=i\eta/2$, 
 therefore we may substitute $\fa(x_1)\to 0$. This results in
\begin{equation}
\label{Hdef2}
  H(x,x_1)=-q_+(x,x_1)-i
\sum_{j=1}^N K(x-u_j) \frac{H(u_j,x_1)}{\fa'(u_j)}, 
\end{equation}
where
\begin{equation}
  \label{qp}
q_+(u,x)=-i(\cot(u-x+i\eta)-\cot(u-x)).
\end{equation}
Introducing the function $F(x,y)=-iH(x,y)/\fa'(x)$ 
we have
\begin{equation}
  F(x,x_1)(i\fa'(x))=-q_+(x,x_1)+
\sum_{j=1}^N K(x-u_j) F(u_j,x_1).
\end{equation}
Specifying to the points $x=u_j$
\begin{equation}
\label{iop}
  F(u_j,x_1)(i\fa'(u_j))=-q_+(u_j,x_1)+
\sum_{k=1}^N K(x-u_k) F(u_k,x_1).
\end{equation}
It is easy to see from \eqref{asol} that
\begin{equation}
  i\fa'(u_j)=L\frac{\sinh(\eta)}{\sin(x-i\eta/2)\sin(x+i\eta/2)}+\sum_{k=1}^N K(u_j-u_k).
\end{equation}
Therefore \eqref{iop} can be written as
\begin{equation}
\label{infected}
  G_{jk} F(u_j,x_1)= q_+(u_j,x_1).
\end{equation}
Evaluating the integral \eqref{Psidef} for the function $\Psi$ leads to
\begin{equation}
\begin{split}
  \Psi(x_1,x_2)
&=2 \sum_{j=1}^N F(u_j,x_1) q(u_j,x_2)-2 \frac{H(x,y)}{1+\fa(x)}+2\frac{q(y,x)}{1+\fa(y)}.
\end{split}
\end{equation}
This can be transformed using equation \eqref{Hdef2} into
\begin{equation}
\Psi(x_1,x_2)=2 \sum_{j=1}^N F(u_j,x_1) q_+(u_j,x_2),
\end{equation}
which is written using \eqref{infected} as
\begin{equation}
\label{XXZpsi}
  \Psi(x_1,x_2)=2 (q_+(u,x_1)\cdot G^{-1}\cdot q_+(u,x_2)).
\end{equation}
Here the multiplication is understood as a summation over the Bethe
roots and $G^{-1}$ is the inverse of the Gaudin matrix. The
derivatives of $\Psi$ around the points $x_{1,2}=i\eta/2$ are given by
\begin{equation}
\label{Psinm}
  \Psi_{n,m}=\partial_{x_1}^n \partial_{x_2}^m \Psi(x_1,x_2)|_{x_{1,2}=i\eta/2}
=2 (q_n \cdot G^{-1} \cdot q_m),
\end{equation}
where we defined
\begin{equation}
  q_j(u)=\partial_{x}^j  q_+(u,ix)|_{x=\eta/2}.
\end{equation}
Note that the functions $q_j(u)$ are the single-particle eigenvalue
functions of the conserved charges. Also, it can be shown that the
first row and column of $\Psi_{n,m}$ are related to 
the conserved charges of the Bethe state in question. Indeed, let $e$ be a vector of length $N$ with all elements
 equal to $1$. It is easy to see from the definition of $G$
that 
\begin{equation}
 L q_0=-G\cdot e.
\end{equation}
It follows that the first row and first column of the matrix $\Psi_{n,m}$ contain the charge densities:
\begin{equation}
  \Psi_{0,n}=\Psi_{n,0}=2 (q_n \cdot G^{-1} \cdot q_0)=
-2\frac{1}{L} (q_n \cdot e)=-2\frac{1}{L}\sum_{j=1}^N q_n(u_j)=-2\frac{Q_n}{L}.
\end{equation}
With similar steps the following algebraic representation can be derived for the function $P$:
\begin{equation}
\label{Pnm}
  \begin{split}
P(x,y)&=2(
-\tilde q_+(u,x)\cdot G^{-1}\cdot q_+(u,y)+
q_+(u,x)\cdot G^{-1}\cdot \tilde q_+(u,y)-\\
&\hspace{4cm}-q_+(u,x)\cdot G^{-1} \cdot \tilde G  \cdot G^{-1}  \cdot q_+(u,y)
),
  \end{split}
\end{equation}
where $\tilde G$ is an other $N\times N$ matrix with elements
\begin{equation}
  \tilde G_{jk}=\tilde K(u_j-u_k)
\end{equation}
and
\begin{equation}
   \tilde q_+(u,x)=-i\cot(u-x+i\eta).
\end{equation}

\subsection{Conjectures for excited states}

The factorization procedure for the correlation functions consists of
the algebraic part and the physical part. Although the factorization for
the excited states has not yet been considered in the literature, it
is very natural to expect that the algebraic 
part of the construction holds also for the excited state of the model
\cite{goehmann-private}, especially in the light of the intermediate
results of \cite{goehmann-kluemper-finite-chain-correlations}.

Regarding the physical part, formulas \eqref{XXZpsi} and \eqref{Pnm} are algebraic
expressions that compute the physical part for the ground state wave function. Using these
expressions, and supplied with the algebraic part, any correlation
function can be expressed as a function of the ground state Bethe
roots.  
Although the resulting formulas were not obtained by a
direct algebraic manipulation of the coordinate space expressions, it
is plausible that for any correlator there is a specific set of
algebraic steps, 
that transforms the ``raw'' real space formulas into the factorized
form, and these would just as well work for the excited states
too. Similarly, for any correlator there is a specific set of 
manipulations that transform the contour integrals into the factorized
form
\cite{boos-goehmann-kluemper-suzuki-fact7,goehmann-kluemper-finite-chain-correlations},
and with a change of contours they would provide the excited state quantities.

Based on the above arguments we formulate the following conjecture:
\begin{conj}
\label{conj:1}
In the XXZ chain the correlation functions of all regular states are
given by the factorized formulas, provided that the
  physical part of the construction is computed via
  \eqref{Psinm}-\eqref{Pnm} using the exact excited state Bethe roots.
\end{conj}
The singular states including the rapidities
$u_j=\pm i\eta/2$ are excluded from the conjecture.
Their true wave function differs from \eqref{xxz-coo-ba}, which becomes
ill-defined. Similarly, the expression \eqref{Psinm} for the building 
blocks $\Psi_{a,b}$ becomes singular. 
Similarly, we excluded the singular states of the root of unity
points that also lead to ill-defined expressions due to the exact $n$-strings.
It is plausible that factorization itself holds for such states, but the
calculation of the physical part needs to be regularized.
These cases will be considered in a future publication.

We performed numerical tests of conjecture \ref{conj:1}. The
methods and some examples of the numerical results are presented in
Appendix \ref{sec:numerics}. In all cases perfect agreement was found.

\section{Excited state correlations of the XXX model}

\label{sec:XXX}

In this section we treat the $SU(2)$-symmetric Heisenberg spin chain,
which is defined through the Hamiltonian 
\begin{equation}
\label{XXXH}
  H=\sum_{j=1}^L (\sigma_j^x\sigma_{j+1}^x+\sigma_j^y\sigma_{j+1}^y+\sigma_j^z\sigma_{j+1}^z-1).
\end{equation}
The coordinate space eigenstates and the Bethe equations can be
obtained either as a scaling limit of the XXZ formulas, or
independently by the Bethe Ansatz of the XXX type. For the sake of
completeness here we summarize the relevant formulas.

The coordinate space wave functions can be written as
\begin{equation}
\label{XXXcoo-ba}
  \phi_N(\{u\}_N|\{y\})=\sum_{\Pe \in S_N} 
\left[\prod_{1\le m < n \le N}\frac{u_{P_m}-u_{P_n}+i}{u_{P_m}-u_{P_n}}\right]
\left[\prod_{l=1}^N
\left(\frac{u_{P_l}+i/2}{u_{P_l}-i/2}\right)^{y_l}\right].
\end{equation}
The Bethe equations take the form
\begin{equation}
 \label{XXXBE}
\left(
\frac{u_j-i/2}{u_j+i/2}
\right)^L \prod_{k\ne j}  
\frac{u_j-u_k+i}{u_j-u_k-i}=1,
\end{equation}
and the energy eigenvalues are
\begin{equation*}
  E=-\sum_j  \frac{2}{u_j^2+1/4}.
\end{equation*}
With the normalization given by \eqref{XXXcoo-ba} the norm of the Bethe 
state is
\begin{equation}
\label{XXXnorm}
  \skalarszorzat{\{u\}_N }{ \{u\}_N}=\prod_j (u_j^2+1/4)
\prod_{j<k} \frac{u_{jk}^2+1}{u_{jk}^2} 
\times \det G,
\end{equation}
where the XXX-type Gaudin matrix $G$ is of the form
\begin{equation}
\label{Gaudin}
  G_{jk}=\delta_{jk}\left(
L \frac{1}{u_j^2+1/4}+\sum_{l=1}^N \varphi(u_{jl})
\right)-\varphi(u_{jk}),
\end{equation}
and
\begin{equation}
  \label{phidef}
\varphi(u)=-\frac{2}{u^2+1}.
\end{equation}

It is known that in the XXX chain the Bethe states are highest weight states with
respect to the $SU(2)$ symmetry and singlet states are obtained when
$N=L/2$ \cite{Faddeev-ABA-intro}. The 
remaining states can be constructed using the global spin lowering operator $S^-$,
which can be embedded naturally into to the Algebraic Bethe Ansatz
framework  \cite{Faddeev-ABA-intro}. In fact, the action $S^-$ is
equivalent to adding a Bethe particle with infinite 
rapidity. However, in the present paper we 
will only consider the highest weight cases.

\subsection{Factorized correlation functions of the XXX chain}

Here we present the known factorized results for the correlators of
the finite XXX chain; we follow the presentation of
\cite{goehmann-kluemper-finite-chain-correlations,boos-goehmann-kluemper-suzuki-fact7,XXXfactorization}. 
It is important that the corresponding results do not follow from a
scaling limit of the factorized XXZ formulas. In fact, the building
blocks $\omega_{a,b}$ and $W_{a,b}$ defined in \eqref{omW} become
singular in the XXX limit and only the special combinations for the
correlators remain finite. In other words, the physical and algebraic
parts of the construction mix with each other. Nevertheless, certain
basic objects such as the auxiliary functions $\fa(x)$, $H(x,y)$ and
the function $\Psi(x,y)$ have simple scaling limits.

The auxiliary function of the XXX model is defined as
\begin{equation}
  \label{adef}
 \log \fa(x)=a_0(x)+\frac{1}{2\pi} 
\int_{\CC} \varphi(x-y) \log(1+\fa(y)) dy,
\end{equation}
where $\varphi(x)$ is given by \eqref{phidef} and $\CC$ is a closed
contour around the real axis, which lies within the strip $|\Im z|<1/2$.
For the finite volume ground state  the source term is
\begin{equation}
  a_0(x)=L\log \frac{x-i/2}{x+i/2}.
\end{equation}
We define the functions
\begin{equation}
\label{omegapsi}
\begin{split}
\Psi^{XXX}(x_1,x_2)&=\frac{1}{\pi} \int_\CC \frac{dy}{1+\fa(y)}
\frac{H(y,x_1)}{(y-x_2)(y-x_2-i)}\\
 \omega(x_1,x_2)&=\frac{1}{2}+\frac{1}{2}((x_1-x_2)^2-1)
 \Psi^{XXX}(ix_1,ix_2),
\end{split}
\end{equation}
where $H$ is the solution to the linear integral equation
\begin{equation}
\label{Hdef}
  H^{XXX}(x,x_1)=-\frac{1}{(x-x_1)(x-x_1-i)}+
\frac{1}{\pi} \int_\CC \frac{dy}{1+\fa(y)} \frac{H^{XXX}(y,x_1)}{1+(x-y)^2}.
\end{equation}
In the previous definitions it is important that the parameters $x_{1,2}$ are assumed to lie
within the contour $C$.

For the finite volume ground state (or in the finite temperature case
with zero magnetic field) all reduced density matrix elements can be
expressed using the functions $\omega$ or $\Psi$ alone.
For the finite volume situation we define
\begin{equation}
 \omega_{n,m}=\partial_{x_1}^n \partial_{x_2}^m \omega(x_1,x_2)|_{x_1,x_2=1/2}\qquad
 \Psi^{XXX}_{n,m}=\partial_{x_1}^n \partial_{x_2}^m \Psi^{XXX}(x_1,x_2)|_{x_1,x_2=i/2}
\end{equation}
Then all ground state correlators can be expressed as a finite
combination of the quantities $\Psi^{XXX}_{n,m}$, for example the simplest $z-z$
correlators read\footnote{In \cite{XXXfactorization}
  the correlators are given in terms of $\omega$, but for convenience
  we present them as a function of $\Psi$. In formula (11) of
  \cite{XXXfactorization} there is a misprint in the case of
  $\vev{\sigma_1^z\sigma_4^z}$: the coefficient of the term
  $(1,0)(3,0)$ is written as -4/27, whereas correctly it is 4/27.}
\begin{align}
\label{sig13}
    \vev{\sigma_{1}^z\sigma_2^z}=&
\frac{1}{3}(1-\Psi^{XXX}_{0,0})\\
   \vev{\sigma_{1}^z\sigma_3^z}
=&\frac{1}{3}(1-4\Psi^{XXX}_{0,0}+\Psi_{1,1}-\frac{1}{2}\Psi^{XXX}_{2,0})\\
\label{sig14}
\begin{split}
  \vev{\sigma_{1}^z\sigma_4^z}=&
\frac{1}{108} (36  +  288 \Psi^{XXX}_{1, 1}   - 15 \Psi^{XXX}_{2, 2} +   10
\Psi^{XXX}_{3, 1}
 +  \Psi^{XXX}_{2, 0}( - 156 + 12 \Psi^{XXX}_{1, 1}  - 6 \Psi^{XXX}_{2, 0})
\\
&+   2 \Psi^{XXX}_{0, 0} (-162  - 42 \Psi^{XXX}_{1, 1}  +   3 \Psi^{XXX}_{2, 2} - 2 \Psi^{XXX}_{3, 1})+\\
&+
\Psi^{XXX}_{1,0}(84 \Psi^{XXX}_{1, 0} - 12 \Psi^{XXX}_{2, 1}+    4 \Psi^{XXX}_{3, 0})).
\end{split}
\end{align}
It is important that in the finite temperature case these formulas
only hold if the magnetization is zero \cite{boos-goehmann-kluemper-suzuki-fact7}. At finite magnetic field the
resulting formulas are more complicated, because they involve additional
objects that were called ``moments'' in
\cite{boos-goehmann-kluemper-suzuki-fact7}. They are defined through
\begin{equation}
\Phi_j(x)=\frac{1}{\pi}  \int_\CC dy \frac{y^{j-1}H(y,x)}{1+\fa(y)}.
\end{equation}
It can be shown that for the ground state in the thermodynamic limit
\begin{equation}
  \lim_{L\to\infty} \Phi_j(x)\equiv \Phi_j^0(x)=(-i\partial_y)^{(j-1)} 
\left.\frac{2e^{iyx}}{1+e^y}\right|_{y=0}.
\end{equation}
The first few cases are
\begin{equation}
  \Phi_1^0(x)=1\qquad
  \Phi_2^0(x)=x+\frac{i}{2}\qquad
  \Phi_3^0(x)=x^2+ix.
\end{equation}
The normalized moments are defined as
\begin{equation}
  \tilde \Phi_j(x)=\Phi_j(x)-\Phi_j^0(x).
\end{equation}
The first normalized moment is special because for the finite volume ground state
(or the finite $T$ case with $h=0$) it vanishes
for arbitrary $x$ \cite{boos-goehmann-kluemper-suzuki-fact7,goehmann-kluemper-finite-chain-correlations}.
It is also useful to introduce the symmetric combinations
\begin{equation}
\label{Deltan}
  \Delta_n(x_1,\dots,x_n)=\frac{\det_n \left(\tilde \Phi_j(x_k)\right)}
{\prod_{1\le j\le k\le n } x_{jk}}.
\end{equation}
It was conjectured in \cite{boos-goehmann-kluemper-suzuki-fact7} that
for $T,h\ne 0$ all local correlators can be expressed using the
functions $\Psi^{XXX}$ and the $\Delta_n$. However, in contrast to the zero
magnetization case it is not known how to compute the algebraic part
for an arbitrary operator. In
\cite{boos-goehmann-kluemper-suzuki-fact7} the reduced density
matrices up to length 3 were computed explicitly, but a general
theory is still missing. An exponential form for the reduced density
matrix is only available for zero magnetization, where all $\Delta_n$
vanish \cite{boos-goehmann-kluemper-suzuki-fact7}. This fact has
 implications also for the excited state correlations.

Returning to the finite volume ground state, all of the previous
integral formulas can be transformed into algebraic expressions, in
the same way as in the XXZ case.  We refrain
from repeating the calculation and just present the final result for
the function $\Psi^{XXX}$:
\begin{equation}
\label{XXXpsi}
  \Psi^{XXX}(x_1,x_2)=2 (q^{XXX}_+(u,x_1)\cdot G^{-1}\cdot q^{XXX}_+(u,x_2)),
\end{equation}
where
\begin{equation}
\label{qdef}
q^{XXX}_+(u,x)=-\frac{1}{(u-x+i)(u-x)}.
\end{equation}
Also, the moments can be expressed as
\begin{equation}
  \label{moments2}
\Phi_j(x)=2
(u^{(j-1)}\cdot G^{-1}\cdot q^{XXX}_+(u,x))+2\frac{x^{j-1}}{1+\fa(x)}.
\end{equation}
Note that \eqref{XXXpsi} has the same structure as the corresponding
formula \eqref{XXZpsi} of the XXZ chain.

\subsection{Excited state correlations}

Here we formulate our conjecture for the excited states of the XXX
model. In this case some care needs to be taken due to the
$SU(2)$-symmetry of the model. 
Both the states and the operators organize themselves into
$SU(2)$-multiplets, and the mean values within each multiplet can be
calculated using the Wigner-Eckart theorem.  
The regular Bethe states are highest weight states, and the finite volume ground state is a
singlet. A priori there is no reason to expect that the factorized formulas for
the ground state should describe the correlations in an
arbitrary $SU(2)$-multiplet. For example the $z-z$ and $x-x$
correlators are typically different. However, the factorized formulas
could hold if the state is a singlet, or if the operator is a
singlet. Regarding the second option it is useful to define the 
$SU(2)$-averaged operators
\begin{equation*}
\bar\ordo=\int_{U\in SU(2)} \mathcal{D}U\ U\ordo U^\dagger,
\end{equation*}
where $\mathcal{D}U$ is the Haar-measure. 
Examples are given by the  operators
\begin{equation}
  \sigma_{1n}\equiv \frac{1}{3}\left(
\sigma_1^x\sigma_n^x+ \sigma_1^y\sigma_n^y +\sigma_1^z\sigma_n^z
\right).
\end{equation}

For the group-invariant operators we formulate the following
conjecture:
\begin{conj}
\label{conj:3}
  For any regular Bethe state of the XXX chain
the mean
  values of the $SU(2)$-invariant operators $\bar\ordo$ are given by
  the known factorized formulas, provided that the 
  physical part of the construction is computed via \eqref{Psinm}
  using the exact excited state Bethe roots.
\end{conj}
This conjecture includes those cases where the Bethe state is a
singlet and the operator $\ordo$ is not, because in singlet states the
mean values of $\ordo$ and $\bar\ordo$ coincide. 
We note that the present situation (namely that relatively simple results hold for group
invariant operators) is analogous to the case of the quantum group invariant
operators of the XXZ chain considered in \cite{bjmst-fact6}.

We tested this conjecture
for the operators $\sigma_{1n}$ for $n=2,3,4$. We performed exact
diagonalization and found perfect agreement on chains with length up
to $L=12$; examples of our data is presented in Appendix
\ref{sec:numerics}. Also, we performed coordinate Bethe Ansatz
calculations for $N=1$ and $N=2$ and arbitrary $L$, and this also
confirms the conjecture. The calculations
are presented in Appendix \ref{sec:app1}.

It is an interesting open question whether some kind of factorization holds for
the mean values of an arbitrary operator $\ordo$ in non-singlet
states. The results of
\cite{boos-goehmann-kluemper-suzuki-fact7,goehmann-kluemper-finite-chain-correlations}
suggest that the multiple integrals can indeed be factorized, 
and the generic case involves the moments too.
In the following section we derive a new result for $\sigma_1^z\sigma_2^z$
which is valid for arbitrary eigenstates, and this result 
confirms the expectations.

\section{The $\Delta\to 1$ limit}

\label{sec:HF}

The goal of this section is to determine the local correlator
$\sigma_1^z\sigma_2^z$ in non-singlet states of the XXX chain. To this
order we employ a careful $\Delta\to 1+$  (or equivalently $\eta\to 0$) limit in finite size. 

As the $\Delta\to 1$ limit is performed from above, the
states organize themselves into  $SU(2)$ multiplets. We follow one
of these states with $N$ particles and assume that the state vector evolves
analytically as a function of $\Delta$.
We apply the Hellmann-Feynman
theorem in the form
\begin{equation}
  L(\bra{\Psi_{XXX}}\sigma_1^z\sigma_2^z\ket{\Psi_{XXX}}-1)=
\lim_{\Delta\to 1} \frac{E_{XXZ}(\Delta)-E_{XXX}}{\Delta-1}.
\end{equation}
At finite $\eta$ the Bethe roots $u_j$ are solutions to the
equations
\begin{equation}
 \label{XXZBE2}
\left(
\frac{\sin(u_j-i\eta/2)}{\sin(u_j+i\eta/2)}
\right)^L \prod_{k\ne j}  
\frac{\sin(u_j-u_k+i\eta)}{\sin(u_j-u_k-i\eta)}=1.
\end{equation}
The energy is given by
\begin{equation}
\label{edf}
  E_{XXZ}(\Delta)=\sum_{j=1}^N -2i \sinh(\eta)\left(
\cot(u_j+i\eta/2)-\cot(u_j-i\eta/2)
\right).
\end{equation}

We assume that the roots $u_j$ scale smoothly into 
the XXX rapidities with the usual behaviour\footnote{The states that
  are not highest weight are obtained when some of the $u_j$ 
don't scale to 0: this results in infinite $x_j$ parameters. However,
here we only consider the highest weight cases.}
\begin{equation}
\label{sc}
  u_j \to \eta x_j,
\end{equation}
such that $x_j$ is a solution of the Bethe equations \eqref{XXXBE}.

At the XXX point the energy becomes
\begin{equation}
  E_{XXX}=\sum_{j=1}^N -2i \left(
\frac{1}{x_j+i/2}-\frac{1}{x_j-i/2}
\right).
\end{equation}
The leading order correction in $\Delta$ is
\begin{equation}
  \Delta-1=\frac{\eta^2}{2}+\dots,
\end{equation}
therefore we need the $\ordo(\eta^2)$ corrections in the energy
\eqref{edf}. The scaling \eqref{sc} gives the correct leading
behaviour, but the first corrections to the
rapidities also need to be calculated. We write
\begin{equation}
  u_j=\eta \tilde u_j,\qquad\tilde u_j=x_j+\eta^2 y_j+\ordo(\eta^4).
\end{equation}
The $y_j$ parameters can be determined from the Bethe equations.
After taking the logarithm of \eqref{XXZBE2} 
we perform the expansions
\begin{equation}
\begin{split}
  \log \frac{\sin(\eta(\tilde u_j-i/2))}{\sin(\eta(\tilde u_j+i/2))}=
&\log \frac{x_j-i/2}{x_j+i/2}+i \eta^2\frac{1}{x_j^2+1/4} y_j
+i\frac{\eta^2}{3}  x_j+\ordo(\eta^4)
\end{split}
\end{equation}
and
\begin{equation}
  \begin{split}
  & \log \frac{\sin(\eta(\tilde u_j-\tilde u_k+i/2))}{\sin(\eta(\tilde
     u_j-\tilde u_k-i/2))}=\\
&\hspace{1cm}
\log \frac{x_j-x_k+i/2}{x_j-x_k-i/2}
-i \eta^2\frac{2}{(x_j-x_k)^2+1} (y_j-y_k)
-2i\frac{\eta^2}{3}  (x_j-x_k)+\ordo(\eta^4).
  \end{split}
\end{equation}
This results in
\begin{equation}
0=
G_{jk}y_k+\frac{1}{3} \left((L-2N)x_j+2\sum_{l} x_l \right),
\end{equation}
which fixes the $y_j$ parameters.

The XXZ energy can be expanded as
\begin{equation}
  \begin{split}
 E_{XXZ}(\Delta)&=
\left(1+\frac{\eta^2}{6}\right)E_{XXX}-\frac{2}{3}\eta^2N-
\eta^2\sum_{j=1}^N 2q_1(x_j) y_j+\ordo(\eta^4)
  \end{split}
\end{equation}
with
\begin{equation}
  q_1(x)=-\frac{2x}{(x^2+1)^2}.
\end{equation}
Putting everything together
\begin{equation}
\label{zz2mindig}
\begin{split}
&  \bra{\Psi_{XXX}}\sigma_1^z\sigma_2^z\ket{\Psi_{XXX}}=\\
&=\frac{L+E_{XXX}}{3L}+\frac{2(L-2N)}{3L}\left[1
+2\sum_{j,k=1}^N q_1(x_j)G^{-1}_{jk} x_k \right]
+\frac{2}{3L}(\sum_j x_j) \sum_{j,k=1}^N 4q_1(x_j) G^{-1}_{jk}.
\end{split}
\end{equation}
This is a new result of the present work. It is interesting to compare
it to equation (29) of \cite{boos-goehmann-kluemper-suzuki-fact7},
which states, that in the finite temperature case with a finite
magnetic field the corresponding correlator can be expressed as
\begin{equation}
\label{hjkl}
  \vev{\sigma_1^z\sigma_2^z}_{T,h}=\frac{2}{3}\Delta_2(i/2,i/2)
+\frac{1}{3} (1-\Psi^{XXX}_{0,0}),
\end{equation}
where $\Delta_2(i/2,i/2)$ is the homogeneous limit of the function defined
in \eqref{Deltan}. 
In Appendix \ref{sec:appzz} it is shown that if we substitute our finite
volume formulas \eqref{Psinm} and \eqref{moments2}  into  \eqref{hjkl}, then we obtain our result
\eqref{zz2mindig}. This is an independent confirmation
of the conjecture that the formulas of
\cite{boos-goehmann-kluemper-suzuki-fact7} with the non-vanishing
moments could work for all excited states of the XXX model.

It is also interesting to specify the result \eqref{zz2mindig}  to singlet
states. The derivation holds also in this case,
but as a result of the $SU(2)$-invariance the $z-z$ correlator is
related directly to the energy, and the correlator is given simply by the
first term of \eqref{zz2mindig}. On the other hand, the second
term vanishes automatically due to $N=L/2$. It follows that
\eqref{zz2mindig} can be valid only if
\begin{equation}
\label{erdekes}
 \left(\sum_j x_j\right) \left(\sum_{j,k=1}^N 4q_1(x_j) G^{-1}_{jk}\right)=0.
\end{equation}
Quite interestingly both factors are zero for all singlet states, and
this can be shown using the following arguments.

The vanishing of the first factor follows from 
the sum rules originally discovered by Baxter in the
context of the XYZ model \cite{BaxterBook}\footnote{The idea of this
  proof was first suggested to us by Andreas Kl\"umper.}. In the XYZ
model the sum of the rapidities in an arbitrary eigenstate with $N=L/2$ is an integer
multiple of $\pi/2$, and this property survives also in the XXZ limit, see for
example equation (17) of \cite{xxz-discrete-symmetries} or (151) of \cite{baxter-completeness}. In the
$\Delta\to 1$ limit the rapidities get rescaled as \eqref{sc}, and if
all the resulting XXX rapidities are finite (ie. the state is really a
singlet), then the only possibility is that this integer multiple of
$\pi/2$ is actually zero.

The vanishing of the second factor follows from the fact, that at zero
magnetization the first moment $\Phi_1(x)$ vanishes for arbitrary $x$ \cite{boos-goehmann-kluemper-suzuki-fact7},
and from \eqref{moments2} we have
\begin{equation}
0=  \partial_x \tilde \Phi_1(x)|_{x\to i/2}=\sum_{j,k=1}^N 4q_1(x_j) G^{-1}_{jk}.
\end{equation}

\section{Conclusions and Outlook}

\label{sec:CO}

In this work we have studied factorized formulas for the excited state
mean values of the XXZ and XXX spin chains. The main idea was that the
known construction for the ground state correlators should give
correct results for the excited states as well, if the physical part is
calculated through certain algebraic expressions, that can be obtained
from the integral representations. Our main conjectures \ref{conj:1} and \ref{conj:3} were tested
using exact diagonalization and real space calculations with low
particle numbers $N=1,2$. The findings can be summarized as follows:
\begin{itemize}
\item In the XXZ case the simple generalization of the ground state
  formulas to excited states works for almost all states and arbitrary
  $\Delta\ne 1$ values. The only exclusions are Bethe states with
  the pair of singular rapidities $\pm i\eta/2$, and the special states of
  the root of unity points $\Delta=\cos(p\pi/q)$. However, we expect
  that the exclusion of these states is a just a technical difficulty, which can 
 be easily circumvented by a proper regularization. 
\item At the XXX point the factorized formulas give correct answers
  for group-invariant operators. This is true for all states except
  those with singular rapidities $\pm i/2$, where regularization is
  needed.
\end{itemize}
We would like to stress that our results can be applied whenever the
algebraic part of the construction is already available. In particular this
means that the distance of the correlator is limited to small
values; in principle the algebraic part could be computed for any
distance, but the resulting expressions become too big and the
calculation becomes unfeasible \cite{kluemper-goehmann-finiteT-review}.

There are several open questions that deserve further
study. First of all, it needs to 
 be shown rigorously whether our main
conjectures  \ref{conj:1} and \ref{conj:3} are correct. 
We have presented evidence that supports the conjectures, but a
rigorous proof would be desirable. 
Also, it needs to be sorted out how to
regularize the factorized formulas to accommodate the states with
singular rapidities. We expect that simple regularization schemes
already available in the literature would solve these problems.

An other,  more ambitious task is to
consider the non-singlet operators in non-singlet
states of the XXX model. The situation is analogous to the case of the finite
temperature correlations with finite magnetic field.
Explicit factorization of multiple integral formulas for this case has been
performed earlier in \cite{boos-goehmann-kluemper-suzuki-fact7}, and it was conjectured that
all short-range correlators can be expressed using the functions
$\Psi^{XXX}$ and $\Delta_n$. Our result \eqref{zz2mindig} about the nearest neighbour $z-z$ correlator has
 the exact same structure as the corresponding formula of
\cite{boos-goehmann-kluemper-suzuki-fact7}. This evidence, together
with the fact that the multiple integral formulas in the finite $T$ and finite size
problems have the same structure
\cite{boos-goehmann-kluemper-suzuki-fact7,goehmann-kluemper-finite-chain-correlations} 
suggest that the generic finite size mean values will take the same
form as in the infinite volume, finite
$T$ problem with finite magnetic field.

Finally, it would be worthwhile to consider the
thermodynamic limit of our finite size formulas. 
In large volumes the summation over the rapidities leads to integrals
over the root densities, and due to the string hypothesis one has to deal
with root densities and other auxiliary functions for all string
types. 
Performing this calculation would establish a bridge to the TBA-like description
of the physical part conjectured in  \cite{sajat-corr}  for $\Delta>1$.
Also, it is important to consider the thermodynamic limit in the
$\Delta<1$ case, because its relevance to quench problems \cite{Dkisebb1,jacopo-michael-hirota}.

We hope to return to these questions in future research.

\vspace{1cm}
{\bf Acknowledgments} 

\bigskip

We would like to thank Frank G\"ohmann, G\'abor Tak\'acs and Jacopo
De Nardis for very useful discussions and for motivating us to finish and
publish this work.

\appendix

\section{Numerical tests}

\label{sec:numerics}

We performed exact diagonalization in order to check our conjectures.
Our procedure included the following steps:
\begin{itemize}
\item We numerically constructed the transfer matrices \eqref{Tdef}
  (and their XXX counterparts)
  for a few arbitrarily chosen rapidity parameters. We exactly
  diagonalized finite sums of transfer matrices. This method has an
  advantage over diagonalizing the Hamiltonian itself, because it
  removes all unwanted degeneracies and it immediately provides the
  Bethe Ansatz states. We considered spin chains up to length $L=12$.
\item For each eigenstate we numerically computed the mean values of
  certain short range correlators. In the XXZ case we considered
the operators $\sigma^a_{1}\sigma^a_n$
  with $a=x,z$ and $n=2,3,4$. On the other hand, in the XXX case we
  chose the averaged operators $\sigma_{1n}$.
\item The Bethe roots of the individual states were found with the method originally developed
  in \cite{fabriziusBetheRoots1}. The idea is to numerically compute
  the transfer matrix eigenvalues for a finite set of rapidities, and
  afterwards use the famous $T-Q$ relations to find the $Q$ function
  and the Bethe roots. 
For  the  formulas relevant to the XXX model we refer the reader to
  \cite{XXXmegoldasok}, which also includes tables of the XXX Bethe roots up to
  $L=12$ (tables with $L>8$ are found in the supplementary material on
  the arxiv). 
\item We computed the predictions for the correlators using the factorized formulas, and
  compared them to the numerics from exact diagonalization.
\end{itemize}

It is important that this method enabled us to treat all excited Bethe
states; even the states with the singular rapidities are found directly.

\bigskip

In the XXZ case we considered both the massive and massless
regimes. It was
observed that Conjecture \ref{conj:1} holds for all states except the singular
ones which  include the special rapidities $\pm i\eta/2$. Tables
\ref{tab:XXZ1}-\ref{tab:XXZrap2} include examples of our numerical
data; here we chose the points $\Delta=2$ and $\Delta=0.7$ with $L=8$ in
both cases. Tables
\ref{tab:XXZ1} and \ref{tab:XXZ2} show the energies, particle
numbers, momentum quantum numbers and the correlations of the first
few states, and the root content of the states is shown in Tables
\ref{tab:XXZrap1} and \ref{tab:XXZrap2}. The numerical errors in the
predictions for the correlators were typically of the order
$10^{-14}-10^{-16}$; examples for the errors are given in table
\ref{tab:XXZhiba1}. For some states a larger error is observed (up to
$\ordo(10^{-6})$), but this is probably related to failure of our
numerical program to accurately resolve degeneracies in the spectrum:
it can be seen from the root content that these states are not parity
invariant, therefore they have a degenerate partner with negated root
content.

In the XXX case it was found that Conjecture \ref{conj:3} indeed holds: the
factorized formulas give correct answer for the singlet operators
$\sigma_{1n}$ for all regular states. 
 Table \ref{tab:XXX1} shows
the energies, particle
numbers, momentum quantum numbers and the correlations in all highest weight
states at $L=8$. Table \ref{tab:XXXrap1} shows the corresponding
rapidities. 

\section{Real space calculations with $N=1$ and $N=2$}

\label{sec:app1}

Here we consider the XXX model and perform real space calculations in the case of low particle
numbers $N=1,2$. We only consider the $SU(2)$ averaged operators 
\begin{equation*}
  \sigma_{1n}=\frac{1}{3}\left[\sigma_{1}^z\sigma_n^z+2(\sigma_1^+\sigma_n^-+\sigma_1^-\sigma_n^+)\right].
\end{equation*}
Throughout the calculations we will use the parametrization
\begin{equation}
  a=e^{ip}=\frac{u+i/2}{u-i/2},
\end{equation}
where $u$ is the Bethe rapidity and $p$ is the one-particle pseudo-momentum.

\subsection{$N=1$}

In the one-particle case
the un-normalized wave function can be written as
\begin{equation}
  \phi(a|y)=a^y.
\end{equation}
The norm is
\begin{equation}
  \skalarszorzat{a}{a}=L.
\end{equation}
A simple direct calculation gives the following value for the correlator:
\begin{equation}
\label{sig141pt}
  \bra{a}\sigma_{1n}\ket{a}=\frac{1}{3}\left[
\frac{L-4}{L}+2\frac{a^{n-1}+a^{-(n-1)}}{L}\right].
\end{equation}
This result has to be compared to the conjectures of Section
\ref{sec:XXX}. 
In the one-particle case the Gaudin matrix is just a scalar:
\begin{equation}
  G=L\frac{1}{u^2+1/4}.
\end{equation}
Then \eqref{XXXpsi} gives simply
\begin{equation}
  \Psi^{XXX}(x,y)=\frac{2(u^2+1/4)}{L((x-i/2-u)^2+1/4)((y-i/2-u)^2+1/4)}.
\end{equation}
It is a straightforward calculation to check that  the factorized
results \eqref{sig13}-\eqref{sig14} indeed reproduce \eqref{sig141pt}
with $n=3$ and $n=4$.

\subsection{$N=2$}

Here we choose the following normalization for the wave function:
\begin{equation}
\label{ujnorma}
  \phi(a,b|x,y)=a^xb^y+a^yb^xS(a,b),\qquad x<y,
\end{equation}
where $S$ is the scattering amplitude which es expressed in the $a$-variables as
\begin{equation*}
  S(a,b)=-\frac{1-2b+ab}{1-2a+ab}.
\end{equation*}
The Bethe equations are
\begin{equation*}
  a^LS(a,b)=b^LS(b,a)=1.
\end{equation*}

We performed the real space calculations using the program Mathematica. As
a warm up we calculated the norm of the wave function
\eqref{ujnorma}. After substituting the Bethe equations we obtained
\begin{equation*}
  \skalarszorzat{a,b}{a,b}=L^2-L
\left(1+\frac{ a S(a,b) -bS(b,a)   }{a-b}\right).
\end{equation*}
It is easy to check that this is equal to the Gaudin determinant
up to the overall normalization differences between \eqref{ujnorma}
and \eqref{XXZ-eloallitas}.

Afterwards we performed the real space calculation of the correlators
$\sigma_{1n}$ and obtained them as rational functions of $a$ and
$b$. After substituting the Bethe equations the results take the form
\eqref{hatez}, where the $C_j$ polynomials only depend on $n$ but not
on $L$\footnote{It is important that in the present calculation the distance $n$ is
simply a parameter. This is in contrast with the multiple integrals of
the ABA method, where the number of the integrals grows with $n$. On
the other hand, the ABA results are valid for arbitrary $N$, whereas the
coordinate BA calculations become increasingly complicated as we
increase $N$.}.
The results are lengthy and we refrain from listing them
here. 

We also calculated the predictions of Conjecture \ref{conj:3}. 
In the present case the Gaudin matrix \eqref{Gaudin} has a simple
structure and the function $\Psi^{XXX}$ is easily calculated using two Bethe
rapidities $u$ and $v$ as
\begin{equation}
\begin{split}
&  \Psi^{XXX}(x,y)=\frac{2}{\det G} \times\\
&\quad \times
  \begin{pmatrix}
    \frac{1}{(u-x+i/2)^2+1/4} \\  \frac{1}{(v-x+i/2)^2+1/4}
  \end{pmatrix}
  \begin{pmatrix}
    \frac{L}{u^2+1/4}-\frac{2}{(u-v)^2+1} &  -\frac{2}{(u-v)^2+1}  \\
-\frac{2}{(u-v)^2+1}  &   \frac{L}{v^2+1/4}-\frac{2}{(u-v)^2+1}
  \end{pmatrix}
 \begin{pmatrix}
    \frac{1}{(u-y+i/2)^2+1/4} \\  \frac{1}{(v-y+i/2)^2+1/4}
  \end{pmatrix}.
\end{split}
\end{equation}
Using the formulas \eqref{sig13}-\eqref{sig14} and making the substitutions
\begin{equation*}
  a=\frac{u+i/2}{u-i/2}\qquad  b=\frac{v+i/2}{v-i/2}
\end{equation*}
the predictions can be compared to real space calculations. 

For both $\sigma_{13}$ and $\sigma_{14}$ we found exact agreement
between the resulting formulas.

\section{The correlator $\vev{\sigma_1^z\sigma_2^z}$ in finite and
  infinite volume}

\label{sec:appzz}

Here we evaluate formula \eqref{hjkl} assuming that the quantities
$\Delta_2(0,0)$ and $\Psi^{XXX}(0,0)$ are given by the finite size formulas
of Section \ref{sec:XXX}. We recall that
\begin{equation*}
\begin{split}
  \Psi^{XXX}_{0,0}=2q_0(u)\cdot G^{-1}\cdot q_0(u),\qquad\qquad
\Phi_j(x)=2 u^{j-1}\cdot G^{-1}\cdot q_+(u-x)+2\frac{x^{j-1}}{1+\fa(x)},
\end{split}
\end{equation*}
and
\begin{equation}
  q_k(u)=(\partial_x)^{k} q_+(u-x)|_{x=i/2},
\end{equation}
and the homogeneous limit of the determinants $\Delta_n$ can be
calculated as
\begin{equation*}
  \Delta_n(i/2,\dots,i/2)=\det\left[\frac{ \tilde \Phi_{j,k}}{(k-1)!}\right], 
\end{equation*}
where
\begin{equation*}
\begin{split}
\tilde \Phi_{j,k}&\equiv \left. (\partial_x)^{k-1} \tilde \Phi_j(x)\right|_{x=i/2}=\\  
&=2 u^{j-1}\cdot G^{-1}\cdot q_{k-1}(u)
+\left.2(\partial_x)^{k-1} x^{j-1}\right|_{x=i/2}
-
\left.\left[ (\partial_x)^{k-1}   (-i\partial_y)^{j-1} 
\frac{2e^{iyx}}{1+e^y}\right]\right|_{y=0,x=i/2}.
\end{split}
\end{equation*}
In the present case
\begin{equation*}
  \Delta_2(0,0)=\tilde\Phi_{1,1}\tilde\Phi_{2,2}-\tilde\Phi_{1,2}\tilde\Phi_{2,1}.
\end{equation*}
For $k=1$ we have
\begin{equation*}
  G^{-1}\cdot q_0(u)=-\frac{1}{L}e,
\end{equation*}
where $e$ is a vector with all elements equal 1. It follows that
\begin{equation*}
  \Psi^{XXX}(0,0)=-\frac{2\sum_j q_0(u_j)}{L}=-\frac{E}{L},\qquad
 \Phi_{1,1}=-\frac{2 N}{L}+2,\qquad
\Phi_{2,1}=-\frac{2\sum_j u_j}{L}. 
\end{equation*}
For the normalized $\tilde\Phi$ quantities we have
\begin{align*}
\tilde \Phi_{1,1}&= -\frac{2 N}{L}+1&
\tilde \Phi_{2,1}& = -\frac{2\sum_j u_j}{L}\\
\tilde \Phi_{1,2}& = 2 e \cdot G^{-1}\cdot q_1(u)&
\tilde \Phi_{2,2}& =  2 u\cdot G^{-1}\cdot q_1(u)  +1.
\end{align*}
Putting everything together formula  \eqref{hjkl} indeed yields \eqref{zz2mindig}.

\addcontentsline{toc}{section}{References}
\providecommand{\href}[2]{#2}\begingroup\raggedright\endgroup




\begin{table}
  \centering
  \begin{tabular}{||c||c||c|c||c|c|c|c|c|c||}
\hline
 & $E$ & $N$ & $J$ & $\vev{\sigma_1^z\sigma_2^z}$ &
$\vev{\sigma_1^z\sigma_3^z}$ & $\vev{\sigma_1^z\sigma_4^z}$ &
$\vev{\sigma_1^x\sigma_2^x}$ & $\vev{\sigma_1^x\sigma_3^x}$ &
$\vev{\sigma_1^x\sigma_4^x}$ \\ 
\hline
\hline
1 & -36.1577 & 4 & -0 & 1.00000 & -0.77381 & 0.55947 & -0.54364 & -0.48605 & 0.13559 \\ \hline 
2 & -35.1227 & 4 & 4 & -1.00000 & -0.87304 & 0.75209 & -0.75785 & -0.32213 & 0.05255 \\ \hline 
3 & -31.9185 & 3 & 4 & -0.00000 & -0.44530 & 0.10801 & 0.14514 & -0.54961 & 0.24317 \\ \hline 
4* & -30.4979 & 4 & 4 & 1.00000 & -0.47190 & -0.03764 & 0.21568 & -0.43421 & -0.08096 \\ \hline 
5 & -29.4276 & 3 & 1 & 0.00000 & -0.46515 & 0.12383 & 0.16438 & -0.37408 & 0.01858 \\ \hline 
6 & -29.1610 & 4 & 1 & -1.00000 & -0.44374 & -0.03196 & 0.18650 & -0.37882 & 0.02717 \\ \hline 
7 & -28.3967 & 4 & 3 & -1.00000 & -0.48454 & -0.00201 & 0.24404 & -0.29025 & 0.14360 \\ \hline 
8 & -28.2969 & 4 & -0 & 1.00000 & -0.61733 & 0.30334 & -0.15418 & -0.15122 & 0.20686 \\ \hline 
9 & -27.4782 & 3 & 3 & -0.00000 & -0.40745 & 0.26602 & -0.21529 & -0.30994 & 0.02627 \\ \hline 
10 & -26.9793 & 3 & 0 & -0.00000 & -0.48056 & 0.36874 & -0.26012 & -0.20565 & 0.22899 \\ \hline 
11 & -26.3751 & 4 & 2 & -1.00000 & -0.40264 & -0.15462 & -0.00260 & -0.24581 & 0.06465 \\ \hline 
12 & -26.2433 & 3 & 2 & 0.00000 & -0.39599 & 0.22330 & -0.18488 & -0.24422 & -0.12448 \\ \hline 
13 & -25.6569 & 4 & 3 & 1.00000 & -0.40636 & -0.08168 & -0.01196 & -0.19720 & -0.18874 \\ \hline 
14 & -25.2867 & 4 & 2 & 1.00000 & -0.41200 & -0.03927 & 0.17598 & -0.16842 & -0.08671 \\ \hline 
15 & -24.7075 & 4 & 2 & 1.00000 & -0.46980 & -0.01595 & 0.20141 & -0.07442 & 0.05612 \\ \hline 
16 & -24.4287 & 3 & 2 & 0.00000 & -0.46749 & 0.11275 & 0.22015 & -0.05931 & -0.03089 \\ \hline 
17 & -24.2925 & 4 & 1 & 1.00000 & -0.48180 & -0.01760 & -0.00060 & -0.03648 & 0.18215 \\ \hline 
18* & -24.0000 & 4 & -0 & -1.00000 & -0.50000 & -0.00000 & -0.00000 & -0.00000 & -0.25000 \\ \hline 
19 & -23.6520 & 3 & 1 & 0.00000 & -0.43058 & 0.22230 & -0.10473 & -0.04767 & 0.06131 \\ \hline 
20 & -23.3698 & 4 & 4 & -1.00000 & -0.42333 & 0.06085 & -0.23535 &
                                                                   -0.03728 & 0.03861 \\ \hline
21 & -23.1266 & 2 & 0 & 0.00000 & 0.00725 & 0.10369 & 0.23868 & -0.45266 & 0.34116 \\ \hline 
22 & -23.0670 & 3 & 4 & -0.00000 & -0.36663 & 0.17000 & -0.00872 & -0.07505 & 0.02696 \\ \hline 
23 & -21.4400 & 3 & 3 & -0.00000 & -0.34113 & -0.03974 & 0.20530 & 0.00113 & -0.13741 \\ \hline 
24 & -21.3810 & 4 & 3 & -1.00000 & -0.35755 & -0.04656 & 0.05623 & 0.02124 & -0.15380 \\ \hline 
25 & -21.2481 & 4 & 1 & -1.00000 & -0.43533 & -0.04543 & 0.22808 & 0.10733 & -0.16686 \\ \hline 
26 & -21.0118 & 2 & 2 & -0.00000 & 0.00472 & 0.09993 & 0.24151 & -0.31796 & -0.00000 \\ \hline 
27* & -20.8791 & 4 & 4 & 1.00000 & -0.31781 & -0.18963 & 0.09706 & 0.01287 & -0.21640 \\ \hline 
28 & -20.2925 & 2 & 1 & -0.00000 & 0.02467 & 0.27299 & 0.20234 & -0.29296 & -0.01285 \\ \hline 
29 & -20.2844 & 4 & 0 & 1.00000 & -0.36483 & -0.19859 & 0.13415 & 0.09705 & 0.02393 \\ \hline 
30* & -20.0000 & 3 & -0 & 0.00000 & -0.00000 & -0.25000 & 0.00000 & -0.25000 & -0.12500 \\ \hline 
31 & -20.0000 & 3 & 0 & -0.00000 & -0.35714 & -0.07143 & 0.21429 & 0.10714 & -0.03571 \\ \hline 
32 & -19.3685 & 4 & 0 & 1.00000 & -0.02885 & -0.22253 & -0.24533 & -0.18168 & -0.07996 \\ \hline 
33 & -19.0707 & 3 & 1 & -0.00000 & -0.03722 & -0.14017 & -0.03592 & -0.15470 & 0.02689 \\ \hline 
34 & -18.9234 & 4 & 1 & -1.00000 & -0.03682 & -0.16601 & -0.26115 & -0.14589 & 0.05193 \\ \hline 
35 & -18.8157 & 3 & 2 & -0.00000 & -0.01470 & -0.24160 & -0.01204 & -0.16129 & -0.15896 \\ \hline 
36 & -18.6357 & 4 & 3 & 1.00000 & -0.06897 & -0.14533 & -0.28570 & -0.09576 & 0.17589 \\ \hline 
37 & -18.5112 & 4 & 2 & -1.00000 & -0.01151 & -0.26087 & -0.25683 & -0.14544 & -0.10876 \\ \hline 
38 & -18.4910 & 3 & 3 & 0.00000 & -0.06635 & -0.09953 & -0.06773 & -0.08934 & 0.17045 \\ \hline 
39 & -18.2825 & 4 & 4 & -1.00000 & -0.16403 & -0.04601 & -0.31077 & 0.02137 & 0.37728 \\ \hline 
40 & -17.9394 & 4 & 2 & 1.00000 & -0.06788 & -0.16993 & -0.19477 & -0.05333 & 0.14711 \\ \hline 
41 & -17.8276 & 3 & 1 & 0.00000 & 0.01651 & -0.22108 & -0.01696 & -0.13074 & -0.17328 \\ \hline 
42 & -17.8109 & 3 & 4 & 0.00000 & -0.14863 & 0.04346 & -0.17412 & 0.03545 & 0.40029 \\ \hline 
  \end{tabular}
  \caption{List of the correlation functions in the first few
    eigenstates of the XXZ model for $\Delta=2$ and $L=8$. 
Whenever degenerate states are connected to each other by space or spin
reflection we only kept one of them in the list.
In the table $N$
    denotes the number of Bethe particles, $J=0\dots (L/2)$ is the overall momentum
  quantum number. 
States marked with a star include the singular rapidities $\pm
i\eta/2$. The factorized formulas correctly reproduce the correlators in all
cases except the singular states.}
  \label{tab:XXZ1}
\end{table}

%
%

\begin{table}
  \centering
  \begin{tabular}{||c||c||c|c|c|c|c|c||}
\hline
 & $E$ & $\vev{\sigma_1^z\sigma_2^z}$ &
$\vev{\sigma_1^z\sigma_3^z}$ & $\vev{\sigma_1^z\sigma_4^z}$ &
$\vev{\sigma_1^x\sigma_2^x}$ & $\vev{\sigma_1^x\sigma_3^x}$ &
$\vev{\sigma_1^x\sigma_4^x}$ \\ 
\hline
    \hline
1 &  -36.1577  & 2.2$\times 10^{-16 }$ & 8.9$\times 10^{-16 }$ & 1.8$\times 10^{-14 }$ & 2.3$\times 10^{-15 }$ & 1.9$\times 10^{-16 }$ & 1.5$\times 10^{-14 }$ \\ \hline 
 2  &  -35.1227  & 8.9$\times 10^{-16 }$ & 4.4$\times 10^{-15 }$ & 5.1$\times 10^{-15 }$ & 1.1$\times 10^{-15 }$ & 4$\times 10^{-16 }$ & 8.4$\times 10^{-15 }$ \\ \hline 
 3  &  -31.9185  & 1.6$\times 10^{-15 }$ & 5.5$\times 10^{-15 }$ & 2.3$\times 10^{-14 }$ & 1.1$\times 10^{-16 }$ & 2.2$\times 10^{-15 }$ & 1.5$\times 10^{-14 }$ \\ \hline 
 4*  &  -30.4979  & NaN  & NaN  & NaN  & NaN  & NaN  & NaN  \\ \hline 
 5  &  -29.4276  & 1$\times 10^{-15 }$ & 3.1$\times 10^{-15 }$ & 3.2$\times 10^{-14 }$ & 1.1$\times 10^{-15 }$ & 4.4$\times 10^{-16 }$ & 2.2$\times 10^{-14 }$ \\ \hline 
 6  &  -29.1610  & 2.1$\times 10^{-15 }$ & 7.1$\times 10^{-16 }$ & 2.2$\times 10^{-14 }$ & 2.4$\times 10^{-15 }$ & 9.2$\times 10^{-15 }$ & 1.7$\times 10^{-14 }$ \\ \hline 
  7 &  -28.3967  & 8.9$\times 10^{-16 }$ & 1.1$\times 10^{-15 }$ & 1.3$\times 10^{-14 }$ & 2.8$\times 10^{-16 }$ & 5.6$\times 10^{-17 }$ & 1$\times 10^{-14 }$ \\ \hline 
  8 &  -28.2969  & 8.9$\times 10^{-16 }$ & 5.6$\times 10^{-16 }$ & 2$\times 10^{-14 }$ & 1.1$\times 10^{-15 }$ & 3.9$\times 10^{-16 }$ & 4.4$\times 10^{-15 }$ \\ \hline 
  9 &  -27.4782  & 2.4$\times 10^{-15 }$ & 7.8$\times 10^{-16 }$ & 1.1$\times 10^{-14 }$ & 5.6$\times 10^{-17 }$ & 1$\times 10^{-15 }$ & 1.1$\times 10^{-14 }$ \\ \hline 
 10  &  -26.9793  & 1.1$\times 10^{-15 }$ & 3.9$\times 10^{-16 }$ & 3.6$\times 10^{-15 }$ & 4.4$\times 10^{-16 }$ & 1.9$\times 10^{-15 }$ & 7.7$\times 10^{-16 }$ \\ \hline 
 11  &  -26.3751  & 5.6$\times 10^{-17 }$ & 1.7$\times 10^{-16 }$ & 1.6$\times 10^{-15 }$ & 3.2$\times 10^{-15 }$ & 2.6$\times 10^{-15 }$ & 8.7$\times 10^{-16 }$ \\ \hline 
  12 &  -26.2433  & 7.8$\times 10^{-16 }$ & 5.4$\times 10^{-15 }$ & 4.9$\times 10^{-15 }$ & 8.6$\times 10^{-16 }$ & 2.8$\times 10^{-16 }$ & 7.8$\times 10^{-15 }$ \\ \hline 
 13  &  -25.6569  & 3.1$\times 10^{-13 }$ & 4.1$\times 10^{-12 }$ & 3.6$\times 10^{-13 }$ & 2.4$\times 10^{-13 }$ & 2.5$\times 10^{-12 }$ & 9$\times 10^{-12 }$ \\ \hline 
 14  &  -25.2867  & 1.1$\times 10^{-15 }$ & 4.2$\times 10^{-14 }$ & 9.6$\times 10^{-14 }$ & 2.2$\times 10^{-15 }$ & 2$\times 10^{-14 }$ & 3.7$\times 10^{-14 }$ \\ \hline 
 15  &  -24.7075  & 1.1$\times 10^{-15 }$ & 3.2$\times 10^{-15 }$ & 3.7$\times 10^{-15 }$ & 1.2$\times 10^{-15 }$ & 2$\times 10^{-15 }$ & 4.3$\times 10^{-15 }$ \\ \hline 
  16 &  -24.4287  & 3.3$\times 10^{-16 }$ & 1.7$\times 10^{-15 }$ & 3$\times 10^{-15 }$ & 8.3$\times 10^{-16 }$ & 5$\times 10^{-16 }$ & 3.7$\times 10^{-15 }$ \\ \hline 
  17 &  -24.2925  & 1.3$\times 10^{-15 }$ & 1.7$\times 10^{-15 }$ & 9$\times 10^{-15 }$ & 1.2$\times 10^{-15 }$ & 1.7$\times 10^{-16 }$ & 7.6$\times 10^{-15 }$ \\ \hline 
  18* &  -24.0000  & NaN  & NaN  & NaN  & NaN  & NaN  & NaN  \\ \hline 
 19  &  -23.6520  & 1.3$\times 10^{-15 }$ & 1.6$\times 10^{-15 }$ & 1.4$\times 10^{-17 }$ & 2$\times 10^{-16 }$ & 4.9$\times 10^{-16 }$ & 6.7$\times 10^{-15 }$ \\ \hline 
  20 &  -23.3698  & 2.8$\times 10^{-14 }$ & 5.8$\times 10^{-13 }$ & 4.9$\times 10^{-14 }$ & 3$\times 10^{-14 }$ & 2.7$\times 10^{-13 }$ & 2.3$\times 10^{-13 }$ \\ \hline 
  21 &  -23.1266  & 5.3$\times 10^{-16 }$ & 6.7$\times 10^{-16 }$ & 1.2$\times 10^{-15 }$ & 4.9$\times 10^{-15 }$ & 4.8$\times 10^{-15 }$ & 4.2$\times 10^{-16 }$ \\ \hline 
  22 &  -23.0670  & 7.8$\times 10^{-16 }$ & 1.4$\times 10^{-15 }$ & 1.5$\times 10^{-14 }$ & 1.6$\times 10^{-15 }$ & 1.2$\times 10^{-15 }$ & 6.7$\times 10^{-15 }$ \\ \hline 
  23 &  -21.4400  & 5$\times 10^{-16 }$ & 1.6$\times 10^{-15 }$ & 1.4$\times 10^{-14 }$ & 3.5$\times 10^{-16 }$ & 5.6$\times 10^{-16 }$ & 7.4$\times 10^{-15 }$ \\ \hline 
  24 &  -21.3810  & 1.8$\times 10^{-15 }$ & 2.2$\times 10^{-15 }$ & 1.5$\times 10^{-14 }$ & 9.7$\times 10^{-16 }$ & 7.8$\times 10^{-16 }$ & 1.8$\times 10^{-14 }$ \\ \hline 
  25 &  -21.2481  & 1.1$\times 10^{-10 }$ & 3.2$\times 10^{-10 }$ & 6.2$\times 10^{-09 }$ & 1.1$\times 10^{-10 }$ & 5$\times 10^{-10 }$ & 1$\times 10^{-08 }$ \\ \hline 
  26 &  -21.0118  & 5.6$\times 10^{-16 }$ & 1$\times 10^{-15 }$ & 1.8$\times 10^{-14 }$ & 4.4$\times 10^{-16 }$ & 1.4$\times 10^{-15 }$ & 1.5$\times 10^{-14 }$ \\ \hline 
  27* &  -20.8791  & NaN  & NaN  & NaN  & NaN  & NaN  & NaN  \\ \hline 
  28 &  -20.2925  & 1.1$\times 10^{-15 }$ & 1.8$\times 10^{-15 }$ & 9.5$\times 10^{-15 }$ & 3.3$\times 10^{-16 }$ & 3.5$\times 10^{-18 }$ & 1$\times 10^{-15 }$ \\ \hline 
  29 &  -20.2844  & 3.9$\times 10^{-16 }$ & 5.6$\times 10^{-17 }$ & 5.1$\times 10^{-15 }$ & 6.7$\times 10^{-16 }$ & 5.6$\times 10^{-16 }$ & 3.3$\times 10^{-15 }$ \\ \hline 
  30* &  -20.0000  & NaN  & NaN  & NaN  & NaN  & NaN  & NaN  \\ \hline 
 31  &  -20.0000  & 1.7$\times 10^{-16 }$ & 3.2$\times 10^{-15 }$ & 6.9$\times 10^{-15 }$ & 7.2$\times 10^{-16 }$ & 6$\times 10^{-16 }$ & 1$\times 10^{-14 }$ \\ \hline 
 32  &  -19.3685  & 4.1$\times 10^{-15 }$ & 6.6$\times 10^{-13 }$ & 1.3$\times 10^{-12 }$ & 2.2$\times 10^{-15 }$ & 3.6$\times 10^{-13 }$ & 4.3$\times 10^{-12 }$ \\ \hline 
  33 &  -19.0707  & 5$\times 10^{-15 }$ & 1$\times 10^{-14 }$ & 2.7$\times 10^{-14 }$ & 2.3$\times 10^{-15 }$ & 1.8$\times 10^{-14 }$ & 4.1$\times 10^{-14 }$ \\ \hline 
  34 &  -18.9234  & 7.4$\times 10^{-16 }$ & 1.5$\times 10^{-15 }$ & 8$\times 10^{-15 }$ & 4.7$\times 10^{-16 }$ & 1.1$\times 10^{-15 }$ & 2.8$\times 10^{-15 }$ \\ \hline 
  35 &  -18.8157  & 1.4$\times 10^{-13 }$ & 8.3$\times 10^{-14 }$ & 3.8$\times 10^{-13 }$ & 1.2$\times 10^{-13 }$ & 2.7$\times 10^{-13 }$ & 4.1$\times 10^{-13 }$ \\ \hline 
  36 &  -18.6357  & 7.5$\times 10^{-16 }$ & 6.9$\times 10^{-15 }$ & 1$\times 10^{-14 }$ & 2.7$\times 10^{-15 }$ & 1.1$\times 10^{-14 }$ & 1.7$\times 10^{-14 }$ \\ \hline 
  37 &  -18.5112  & 3.3$\times 10^{-15 }$ & 2.1$\times 10^{-14 }$ & 3.1$\times 10^{-14 }$ & 3.8$\times 10^{-15 }$ & 3.1$\times 10^{-14 }$ & 1.4$\times 10^{-14 }$ \\ \hline 
  38 &  -18.4910  & 6.5$\times 10^{-16 }$ & 2.5$\times 10^{-15 }$ & 9.5$\times 10^{-15 }$ & 1.7$\times 10^{-15 }$ & 1.5$\times 10^{-15 }$ & 8.3$\times 10^{-15 }$ \\ \hline 
  39 &  -18.2825  & 2.4$\times 10^{-15 }$ & 6.7$\times 10^{-15 }$ & 1$\times 10^{-14 }$ & 4.5$\times 10^{-15 }$ & 7.8$\times 10^{-15 }$ & 6.4$\times 10^{-15 }$ \\ \hline 
  40 &  -17.9394  & 7.9$\times 10^{-16 }$ & 4.7$\times 10^{-16 }$ & 1.7$\times 10^{-14 }$ & 1.3$\times 10^{-15 }$ & 1.1$\times 10^{-16 }$ & 1.6$\times 10^{-14 }$ \\ \hline 
  41 &  -17.8276  & 8.3$\times 10^{-09 }$ & 4.2$\times 10^{-08 }$ & 2.1$\times 10^{-06 }$ & 8.5$\times 10^{-09 }$ & 4.7$\times 10^{-08 }$ & 1.8$\times 10^{-06 }$ \\ \hline 
42     &  -17.8109  & 1.9$\times 10^{-16 }$ & 8.9$\times 10^{-16 }$ & 9.4$\times 10^{-15 }$ & 2.6$\times 10^{-15 }$ & 1.9$\times 10^{-15 }$ & 5.1$\times 10^{-15 }$ \\ \hline 
  \end{tabular}
  \caption{List of the numerical errors for the calculation of correlation functions in the first few
    eigenstates of the XXZ model for $\Delta=2$ and $L=8$. 
Whenever degenerate states are connected to each other by space or spin
reflection we only kept one of them in the list.
States marked with a star include the singular rapidities $\pm
i\eta/2$; in these cases the factorized correlations are not computed,
because the corresponding expressions are ill defined and need regularization.}
  \label{tab:XXZhiba1}
\end{table}

\begin{table}
  \centering
  \begin{tabular}{||c||c|c|c|c||}
\hline
1  & -0.16931 & 0.16931 & -0.67692 & 0.67692\\ \hline 
2  & 0 & 0.36517 & -0.36517 & -1.57080\\ \hline 
3  & 0 & -0.32582 & 0.32582 & \\ \hline 
4*  & 0.18374 & -0.18374 & -0.65848i & 0.65848i\\ \hline 
5  & 0.08677 & -0.22094 & -0.68609 & \\ \hline 
6  & 0.03422 & -0.33383 & -0.63559+0.65091i & -0.63559-0.65091i\\ \hline 
7  & 0.01707 & 0.37440 & -0.98113+0.61044i & -0.98113-0.61044i\\ \hline 
8  & -0.15916 & 0.15916 & -1.57080+0.76282i & -1.57080-0.76282i\\ \hline 
9  & 0.03724 & 0.36952 & -0.78636 & \\ \hline 
10  & 0.14958 & -0.14958 & -1.57080 & \\ \hline 
11  & -0.32884 & 0.38712 & -0.81454+0.63423i & -0.81454-0.63423i\\ \hline 
12  & -0.24837 & 0.39897 & -0.73478 & \\ \hline 
13  & 0.16131 & 0.27757-0.65850i & 0.27757+0.65850i & -0.71645\\ \hline 
14  & -0.18828 & -0.70030 & 0.44429-0.65896i & 0.44429+0.65896i\\ \hline 
15  & 0.13162 & 0.57017 & 1.21990+0.75972i & 1.21990-0.75972i\\ \hline 
16  & 0.10182 & 0.46175 & 1.39682 & \\ \hline 
17  & -0.17285 & 0.57575 & 1.36934+0.76935i & 1.36934-0.76935i\\ \hline 
18*  & 0 & -0.65848i & 0.65848i & -1.57080\\ \hline 
19  & -0.17459 & 0.49080 & 1.47126 & \\ \hline 
20  & 0 & 1.57080 & -1.57080-0.35091i & -1.57080+0.35091i\\ \hline 
21  & 0.13785 & -0.13785 &  & \\ \hline 
22  & 0 & -0.87010 & 0.87010 & \\ \hline 
23  & -0.28888 & -0.81653 & 0.91977 & \\ \hline 
24  & -0.32729 & 1.26917 & -1.25634+0.47792i & -1.25634-0.47792i\\ \hline 
25  & 0.38794 & -0.17943+0.65848i & -0.17943-0.65848i & 1.54172\\ \hline 
26  & 0.08798 & 0.40130 &  & \\ \hline 
27*  & -0.65848i & 0.65848i & -0.71693 & 0.71693\\ \hline 
28  & -0.16404 & 0.43048 &  & \\ \hline 
29  & -0.58190 & 0.58190 & 1.57080-0.78528i & 1.57080+0.78528i\\ \hline 
30*  & 0 & 0.65848i & -0.65848i & \\ \hline 
31  & 0.52360 & -0.52360 & -1.57080 & \\ \hline 
32  & 0.04168 & -0.04168 & 1.33652i & -1.33652i\\ \hline 
33  & -0.11381 & 0.52304-0.65895i & 0.52304+0.65895i & \\ \hline 
34  & -0.10163 & 0.54749 & 0.56247-1.32488i & 0.56247+1.32488i\\ \hline 
35  & 0.18614 & 0.41430+0.65859i & 0.41430-0.65859i & \\ \hline 
36  & 0.08631 & 0.98431 & 1.03549+1.36050i & 1.03549-1.36050i\\ \hline 
37  & 0.18214 & 0.45323 & 0.46771+1.33824i & 0.46771-1.33824i\\ \hline 
38  & 0.07731 & 0.98921-0.64251i & 0.98921+0.64251i & \\ \hline 
39  & 0 & -1.57080 & -1.57080+1.41573i & -1.57080-1.41573i\\ \hline 
40  & -0.19570 & 1.07228 & 1.13250-1.35314i & 1.13250+1.35314i\\ \hline 
41  & 0.33452 & -0.12946+0.65848i & -0.12946-0.65848i & \\ \hline 
42  & 0 & -1.57080+0.69453i & 1.57080-0.69453i & \\ \hline 
  \end{tabular}
  \caption{Bethe root content in the first few states at $\Delta=2$, $L=8$. 
Whenever degenerate states are connected to each other by space or spin
reflection we only kept one of them in the list.
States
  marked with a star include the singular rapidities $\pm i\eta/2$. }
  \label{tab:XXZrap1}
\end{table}

\begin{table}
  \centering
  \begin{tabular}{||c||c||c|c||c|c|c|c|c|c||}
\hline
 & $E$ & $N$ & $J$ & $\vev{\sigma_1^z\sigma_2^z}$ &
$\vev{\sigma_1^z\sigma_3^z}$ & $\vev{\sigma_1^z\sigma_4^z}$ &
$\vev{\sigma_1^x\sigma_2^x}$ & $\vev{\sigma_1^x\sigma_3^x}$ &
$\vev{\sigma_1^x\sigma_4^x}$ \\ 
\hline
\hline
1 & -18.8078 & 4 & 0 & -0.55523 & 0.17411 & -0.17857 & -0.63116 & 0.30760 & -0.29264 \\ \hline 
2 & -17.1789 & 3 & 4 & -0.37804 & 0.05519 & 0.09571 & -0.59137 & 0.34351 & -0.26778 \\ \hline 
3 & -16.3463 & 4 & 4 & -0.70565 & 0.46055 & -0.50445 & -0.42467 & 0.07272 & 0.15524 \\ \hline 
4* & -15.3317 & 4 & 4 & -0.43852 & -0.07926 & 0.18411 & -0.45475 & -0.05900 & 0.20029 \\ \hline 
5 & -14.5716 & 4 & 1 & -0.34075 & -0.08169 & 0.08617 & -0.44147 & 0.12485 & 0.10379 \\ \hline 
6 & -14.4385 & 3 & 1 & -0.40282 & 0.07618 & 0.11199 & -0.41142 & 0.01125 & 0.15510 \\ \hline 
7 & -13.5074 & 3 & 3 & -0.25626 & 0.10481 & -0.15772 & -0.40452 & 0.09997 & 0.01195 \\ \hline 
8 & -13.0793 & 4 & 3 & -0.45086 & -0.00254 & 0.21038 & -0.30966 & 0.15167 & -0.16087 \\ \hline 
9 & -12.8462 & 2 & 0 & 0.01812 & 0.11432 & 0.22837 & -0.45923 & 0.36597 & -0.27813 \\ \hline 
10 & -12.3510 & 3 & 2 & -0.25615 & 0.07089 & -0.13928 & -0.33228 & -0.05083 & -0.07248 \\ \hline 
11 & -12.2647 & 4 & 2 & -0.29826 & -0.29478 & -0.01439 & -0.31215 & 0.13104 & -0.17247 \\ \hline 
12 & -11.8077 & 4 & 3 & -0.23238 & -0.19759 & -0.07003 & -0.30665 & -0.14760 & -0.03699 \\ \hline 
13 & -11.7284 & 3 & 0 & -0.44022 & 0.34938 & -0.26275 & -0.22895 & 0.23119 & -0.00485 \\ \hline 
14 & -11.5032 & 4 & 2 & -0.20335 & -0.11094 & -0.04214 & -0.29777 & -0.07649 & 0.05979 \\ \hline 
15 & -11.4450 & 4 & 0 & -0.50248 & 0.37331 & -0.48284 & -0.18945 & 0.18769 & 0.10722 \\ \hline 
16 & -10.6975 & 2 & 2 & 0.01426 & 0.11117 & 0.23182 & -0.32358 & -0.00000 & 0.18678 \\ \hline 
17 & -10.3143 & 2 & 1 & 0.06514 & 0.26935 & 0.16551 & -0.31745 & 0.03794 & 0.03011 \\ \hline 
18 & -10.0405 & 3 & 2 & -0.19673 & -0.09617 & 0.12150 & -0.20867 & 0.08873 & 0.19586 \\ \hline 
19 & -10.0132 & 3 & 4 & -0.13796 & 0.00401 & 0.09603 & -0.22754 & -0.00317 & -0.09340 \\ \hline 
20 & -9.9144 & 4 & 4 & -0.19588 & -0.02394 & -0.37565 & -0.20109 & 0.06464 & 0.04398 \\ \hline 
21 & -9.6574 & 4 & 2 & -0.40732 & -0.04904 & 0.12664 & -0.11103 & 0.07340 & 0.21892 \\ \hline 
22* & -9.6000 & 3 & 0 & -0.00000 & -0.25000 & -0.00000 & -0.25000 & -0.12500 & 0.00000 \\ \hline 
23 & -9.5110 & 3 & 1 & -0.24273 & -0.07193 & 0.10480 & -0.15949 & 0.18795 & 0.05760 \\ \hline 
24 & -9.1958 & 4 & 1 & -0.37899 & -0.10244 & -0.01856 & -0.09209 & 0.18674 & 0.14175 \\ \hline 
25 & -8.8184 & 4 & 3 & -0.06300 & -0.13441 & -0.24467 & -0.17910 & -0.14975 & 0.12408 \\ \hline 
26 & -8.7270 & 3 & 3 & -0.10430 & -0.21899 & 0.08270 & -0.15893 & -0.14497 & -0.01306 \\ \hline 
27* & -8.4253 & 4 & 4 & -0.05373 & -0.22139 & -0.22437 & -0.15778 & -0.25000 & -0.07695 \\ \hline 
28* & -8.4000 & 4 & 0 & -0.50000 & -0.00000 & 0.00000 & -0.00000 & -0.25000 & 0.00000 \\ \hline 
29 & -8.0873 & 4 & 0 & -0.24214 & -0.05748 & -0.21013 & -0.07071 &
                                                                    -0.18598  & -0.20935 \\ \hline 
30 & -7.9465 & 4 & 1 & -0.07037 & -0.06535 & -0.25547 & -0.12203 & -0.00078 & -0.09941 \\ \hline 
31 & -7.9343 & 4 & 2 & -0.02443 & -0.30708 & -0.24022 & -0.13734 & -0.15321 & 0.03579 \\ \hline 
32 & -7.7855 & 3 & 1 & 0.06800 & -0.18074 & -0.06856 & -0.16039 & -0.14461 & 0.07699 \\ \hline 
33 & -7.7574 & 3 & 1 & -0.17221 & 0.08450 & -0.15357 & -0.07457 & -0.07944 & -0.15130 \\ \hline 
34 & -7.6784 & 2 & 0 & 0.13829 & 0.20570 & 0.01856 & -0.17830 & -0.21003 & 0.08810 \\ \hline 
35 & -7.5238 & 3 & 2 & -0.25121 & -0.05371 & 0.05552 & -0.03231 & -0.23776 & -0.12077 \\ \hline 
36 & -7.4061 & 2 & 3 & 0.03020 & 0.27341 & 0.19640 & -0.12345 & 0.00431 & -0.13014 \\ \hline 
37 & -7.0942 & 3 & 0 & -0.06867 & -0.23635 & -0.02045 & -0.06936 & 0.00256 & 0.07524 \\ \hline 
38 & -7.0588 & 3 & 3 & -0.11784 & 0.05867 & -0.19845 & -0.04993 & 0.12789 & 0.04083 \\ \hline 
39 & -6.9067 & 4 & 3 & -0.20247 & -0.03306 & -0.26446 & -0.01080 & 0.18554 & 0.06709 \\ \hline 
40 & -6.8721 & 4 & 1 & -0.29782 & -0.15369 & 0.18784 & 0.02473 & -0.21028 & -0.02506 \\ \hline 
41 & -6.8651 & 2 & 2 & 0.11190 & 0.23114 & 0.01427 & -0.11823 & -0.00000 & -0.05325 \\ \hline 
42 & -6.8000 & 1 & 4 & 0.50000 & 0.50000 & 0.50000 & -0.25000 & 0.25000 & -0.25000 \\ \hline 
  \end{tabular}
  \caption{
List of the correlation functions in the first few
    eigenstates of the XXZ model for $\Delta=0.7$ and $L=8$. 
Whenever degenerate states are connected to each other by space or spin
reflection we only kept one of them in the list.
In the table $N$
    denotes the number of Bethe particles, $J=0\dots (L/2)$ is the overall momentum
  quantum number. 
States marked with a star include the singular rapidities
$\pm i\eta/2$. The factorized formulas correctly reproduce the correlators in all
cases except the singular states.
}
  \label{tab:XXZ2}
\end{table}

\begin{table}
  \centering
  \begin{tabular}{||c||c|c|c|c||}
\hline
1  & 0.10290i & -0.10290i & -0.41679i & 0.41679i\\ \hline 
2  & 0 & -0.21412i & 0.21412i & \\ \hline 
3  & 0 & -0.19828i & 0.19828i & -1.57080\\ \hline 
4*  & -0.11284i & 0.11284i & 0.39770 & -0.39770\\ \hline 
5  & 0.05523i & -0.13596i & -0.45671i & 1.57080+0.53745i\\ \hline 
6  & 0.03604i & -0.16484i & -0.57849i & \\ \hline 
7  & 0.01595i & 0.22953i & -0.62652i & \\ \hline 
8  & 0.02501i & 0.22723i & -0.50508i & -1.57080+0.25284i\\ \hline 
9  & 0.09368i & -0.09368i &  & \\ \hline 
10  & -0.17550i & 0.24105i & -0.59845i & \\ \hline 
11  & -0.15054i & 0.24671i & -0.47828i & -1.57080+0.38211i\\ \hline 
12  & 0.09671i & -0.39772+0.17721i & 0.39772+0.17721i & -0.45113i\\ \hline 
13  & 0.08816i & -0.08816i & 1.57080 & \\ \hline 
14  & -0.11623i & -0.44009i & -0.39811+0.27816i & 0.39811+0.27816i\\ \hline 
15  & -0.08831i & 0.08831i & -1.57080-0.67395i & 1.57080+0.67395i\\ \hline 
16  & 0.07157i & 0.30728i &  & \\ \hline 
17  & -0.10461i & 0.32066i &  & \\ \hline 
18  & 0.05874i & 0.27040i & -1.57080-0.20712i & \\ \hline 
19  & 0 & 0.64966i & -0.64966i & \\ \hline 
20  & 0 & 0.54135i & -0.54135i & -1.57080\\ \hline 
21  & 0.06005i & 0.26612i & 1.57080+0.46830i & -1.57080-0.79448i\\ \hline 
22*  & 0 & -0.39770 & 0.39770 & \\ \hline 
23  & -0.10375i & 0.28673i & 1.57080-0.11375i & \\ \hline 
24  & -0.10232i & 0.28371i & -1.57080+0.55219i & 1.57080-0.73359i\\ \hline 
25  & -0.17362i & -0.51089i & 0.56888i & -1.57080+0.11563i\\ \hline 
26  & -0.18992i & -0.62127i & 0.66947i & \\ \hline 
27*  & 0.39770 & -0.39770 & -0.44618i & 0.44618i\\ \hline 
28*  & 0 & -0.39770 & 0.39770 & -1.57080\\ \hline 
29  & 0.02845i & -0.02845i & 0.81043 & -0.81043\\ \hline 
30  & -0.07634i & -0.39925+0.36643i & 0.39925+0.36643i & 1.57080-0.65652i\\ \hline 
31  & 0.11033i & -0.39820+0.29906i & 0.39820+0.29906i & -1.57080-0.70845i\\ \hline 
32  & 0.22711i & -0.39770-0.09840i & 0.39770-0.09840i & \\ \hline 
33  & -0.06572i & 0.43479+0.65780i & -0.43479+0.65780i & \\ \hline 
34  & -0.33388i & 0.33388i &  & \\ \hline 
35  & 0.11339i & -0.42937+0.61132i & 0.42937+0.61132i & \\ \hline 
36  & 0.05033i & 0.96789i &  & \\ \hline 
37  & 0.30501i & -0.30501i & -1.57080 & \\ \hline 
38  & 0.03844i & 0.70050i & 1.57080-0.30395i & \\ \hline 
39  & 0.03543i & 0.59541i & 1.57080+0.23043i & -1.57080-0.86128i\\ \hline 
40  & 0.20450i & 0.39770-0.07752i & -0.39770-0.07752i & -1.57080-0.04947i\\ \hline 
41  & -0.12135i & 0.99398i &  & \\ \hline 
42  & 0 &  &  & \\ \hline 
  \end{tabular}
  \caption{
Bethe root content in the first few states at $\Delta=0.7$,
$L=8$. Note that here the ground state rapidities are all purely
imaginary and in the excited states the strings are centered around the imaginary axis; this
is simply a result of our intentions to apply the same conventions for
both the 
$\Delta<1$ and $\Delta>1$ regimes, as explained in the main text.
Whenever degenerate states are connected to each other by space or spin
reflection we only kept one of them in the list.
States
  marked with a star include the singular rapidities $\pm i\eta/2$.
}
  \label{tab:XXZrap2}
\end{table}

\begin{table}
  \centering
  \begin{tabular}{||c||c||c|c||c|c|c||}
\hline
 & $E$ & $N$ & $J$ & 
$\vev{\sigma_{12}}$ & $\vev{\sigma_{13}}$ &  $\vev{\sigma_{14}}$  \\
\hline
\hline
1 & -22.60437 & 4 & 0 & -0.60852 & 0.26104 & -0.25194\\ \hline 
2 & -20.51368 & 3 & 4 & -0.52140 & 0.23367 & -0.12039\\ \hline 
3* & -18.79851 & 4 & 4 & -0.44994 & -0.06564 & 0.19466\\ \hline 
4 & -17.83495 & 3 & 1 & -0.40979 & 0.04117 & 0.12669\\ \hline 
5 & -16.58059 & 3 & 3 & -0.35752 & 0.10082 & -0.04167\\ \hline 
6 & -15.41855 & 3 & 2 & -0.30911 & -0.01099 & -0.10432\\ \hline 
7 & -15.20775 & 2 & 0 & -0.30032 & 0.27600 & -0.09850\\ \hline 
8 & -14.82843 & 4 & 3 & -0.28452 & -0.16667 & -0.04882\\ \hline 
9 & -14.47214 & 4 & 2 & -0.26967 & -0.09213 & 0.02847\\ \hline 
10 & -13.06814 & 2 & 2 & -0.21117 & 0.03583 & 0.19537\\ \hline 
11 & -12.80656 & 3 & 4 & -0.20027 & 0.00522 & -0.05067\\ \hline 
12 & -12.57649 & 2 & 1 & -0.19069 & 0.10541 & 0.08528\\ \hline 
13* & -12.00000 & 3 & 0 & -0.16667 & -0.16667 & 0\\ \hline 
14 & -11.43569 & 3 & 3 & -0.14315 & -0.16631 & 0.01031\\ \hline 
15* & -11.04351 & 4 & 4 & -0.12681 & -0.24628 & -0.13355\\ \hline 
16 & -10.90444 & 4 & 0 & -0.12102 & -0.16067 & -0.21444\\ \hline 
17 & -10.51351 & 3 & 1 & -0.10473 & -0.02145 & -0.16411\\ \hline 
18 & -10.38787 & 3 & 2 & -0.09949 & -0.19118 & -0.07307\\ \hline 
19 & -10.05073 & 3 & 1 & -0.08545 & -0.16809 & 0.03510\\ \hline 
20 & -9.78017 & 2 & 0 & -0.07417 & -0.07899 & 0.05466\\ \hline 
21 & -9.74806 & 2 & 3 & -0.07284 & 0.10541 & -0.03257\\ \hline 
22 & -9.17157 & 4 & 1 & -0.04882 & -0.16667 & -0.28452\\ \hline 
23 & -9.03461 & 2 & 2 & -0.04311 & 0.08643 & -0.02556\\ \hline 
24 & -8.00000 & 2 & 4 & 0 & -0.16667 & 0.08333\\ \hline 
25 & -8.00000 & 1 & 4 & 0 & 0.33333 & 0\\ \hline 
26 & -8.00000 & 3 & 0 & 0 & -0.16667 & -0.08333\\ \hline 
27 & -6.93819 & 3 & 3 & 0.04424 & -0.24910 & -0.09301\\ \hline 
28 & -6.82843 & 1 & 3 & 0.04882 & 0.16667 & 0.28452\\ \hline 
29 & -6.49119 & 4 & 0 & 0.06287 & -0.43370 & -0.20029\\ \hline 
30 & -6.25194 & 2 & 1 & 0.07284 & -0.10541 & 0.03257\\ \hline 
31 & -6.19358 & 3 & 2 & 0.07527 & -0.29784 & -0.15595\\ \hline 
32 & -5.52786 & 4 & 2 & 0.10301 & -0.24120 & -0.19514\\ \hline 
33 & -4.42923 & 3 & 1 & 0.14878 & -0.18496 & -0.21317\\ \hline 
34* & -4.00000 & 2 & 4 & 0.16667 & -0.16667 & 0\\ \hline 
35 & -4.00000 & 1 & 2 & 0.16667 & 0 & 0.16667\\ \hline 
36 & -3.42351 & 2 & 3 & 0.19069 & -0.10541 & -0.08528\\ \hline 
37 & -3.01208 & 2 & 0 & 0.20783 & -0.03034 & -0.12283\\ \hline 
38 & -2.67977 & 3 & 4 & 0.22168 & -0.07222 & -0.32894\\ \hline 
39 & -2.21710 & 3 & 3 & 0.24095 & -0.01875 & -0.32682\\ \hline 
40* & -2.15798 & 4 & 4 & 0.24342 & -0.02141 & -0.39445\\ \hline 
41 & -1.89725 & 2 & 2 & 0.25428 & 0.04441 & -0.16981\\ \hline 
42 & -1.17157 & 1 & 1 & 0.28452 & 0.16667 & 0.04882\\ \hline 
  \end{tabular}
  \caption{
List of the correlation functions in the highest weight states of the XXX model for $L=8$. 
Whenever degenerate states are connected to each other by space reflection we only kept one of them in the list.
In the table $N$
    denotes the number of Bethe particles, $J=0\dots (L/2)$ is the overall momentum
  quantum number. 
States marked with a star include the singular rapidities
$\pm i/2$. 
The factorized formulas correctly reproduce the correlators in all
cases except the singular states.
}
  \label{tab:XXX1}
\end{table}

\begin{table}
  \centering
  \begin{tabular}{||c||c|c|c|c||}
\hline
1  & -0.12947 & 0.12947 & 0.52501 & -0.52501\\ \hline 
2  & 0 & -0.26391 & 0.26391 & \\ \hline 
3*  & -0.14247 & 0.14247 & 0.50000i & -0.50000i\\ \hline 
4  & 0.05372 & -0.19358 & -0.65085 & \\ \hline 
5  & 0.02382 & 0.28883 & -0.72050 & \\ \hline 
6  & -0.20971 & 0.30632 & -0.68166 & \\ \hline 
7  & -0.11412 & 0.11412 &  & \\ \hline 
8  & 0.12119 & 0.22521+0.50003i & 0.22521-0.50003i & -0.57161\\ \hline 
9  & -0.14701 & -0.55707 & 0.35204-0.50056i & 0.35204+0.50056i\\ \hline 
10  & 0.08200 & 0.35910 &  & \\ \hline 
11  & 0 & 0.76302 & -0.76302 & \\ \hline 
12  & -0.13044 & 0.37844 &  & \\ \hline 
13*  & 0 & -0.50000i & 0.50000i & \\ \hline 
14  & -0.23264 & -0.72324 & 0.79382 & \\ \hline 
15*  & 0.50000i & -0.50000i & -0.56383 & 0.56383\\ \hline 
16  & -0.04131 & 0.04131 & -1.02571i & 1.02571i\\ \hline 
17  & -0.08884 & 0.62094-0.51103i & 0.62094+0.51103i & \\ \hline 
18  & 0.13981 & 0.55039-0.50687i & 0.55039+0.50687i & \\ \hline 
19  & 0.27866 & -0.11627+0.50000i & -0.11627-0.50000i & \\ \hline 
20  & 0.39874 & -0.39874 &  & \\ \hline 
21  & 0.05396 & 0.91480 &  & \\ \hline 
22  & -0.08379 & 0.24433 & -0.08027+1.00559i & -0.08027-1.00559i\\ \hline 
23  & -0.15507 & 0.94957 &  & \\ \hline 
24  & 0.28868 & 0.86603 &  & \\ \hline 
25  & 0 &  &  & \\ \hline 
26  & 0.34781 & -0.67391-0.51443i & -0.67391+0.51443i & \\ \hline 
27  & 0.31492+0.50018i & 0.31492-0.50018i & 0.59567 & \\ \hline 
28  & -0.20711 &  &  & \\ \hline 
29  & 0.46326-0.50229i & -0.46326+0.50229i & -0.46326-0.50229i & 0.46326+0.50229i\\ \hline 
30  & -0.42841 & 0.98514 &  & \\ \hline 
31  & -0.21341-0.49999i & -0.21341+0.49999i & 0.77127 & \\ \hline 
32  & -0.22056 & 0.66912 & -0.22428+1.00225i & -0.22428-1.00225i\\ \hline 
33  & 0.86575 & -0.74066-0.51922i & -0.74066+0.51922i & \\ \hline 
34*  & 0.50000i & -0.50000i &  & \\ \hline 
35  & -0.50000 &  &  & \\ \hline 
36  & -0.41534+0.49953i & -0.41534-0.49953i &  & \\ \hline 
37  & -1.03826 & 1.03826 &  & \\ \hline 
38  & 0 & 1.00092i & -1.00092i & \\ \hline 
39  & -0.63120 & -0.61576-0.98815i & -0.61576+0.98815i & \\ \hline 
40*  & 0.50000i & -0.50000i & 1.55613i & -1.55613i\\ \hline 
41  & -0.95114-0.54450i & -0.95114+0.54450i &  & \\ \hline 
42  & -1.20711 &  &  & \\ \hline 
  \end{tabular}
  \caption{
Bethe root content in the heighest weight states of the XXX model for
$L=8$. 
Whenever degenerate states are connected to each other by space
reflection we only kept one of them in the list.
The singular states including rapidities $\pm i/2$ are denoted by a star. 
}
  \label{tab:XXXrap1}
\end{table}

\end{document}